\documentclass[12pt,preprint]{aastex6}
\usepackage{CJK}
\usepackage{color}

\usepackage{natbib}
\usepackage{hyperref}
\usepackage{cleveref}

\shorttitle{High-density Symmetry Energy from Neutron Stars}
\shortauthors{Xie and Li}
\begin{document}
\title{\bf Bayesian Inference of High-density Nuclear Symmetry Energy\\ from Radii of Canonical Neutron Stars}
\author{Wen-Jie Xie\altaffilmark{1,2} and Bao-An Li\altaffilmark{1}$^{*}$}
\altaffiltext{1}{Department of Physics and Astronomy, Texas A$\&$M University-Commerce, Commerce, TX 75429, USA}
\altaffiltext{2}{Department of Physics, Yuncheng University, Yuncheng 044000, China\\
\noindent{$^{*}$Corresponding author: Bao-An.Li@Tamuc.edu}}

\begin{abstract}
The radius $R_{1.4}$ of neutron stars (NSs) with a mass of 1.4 M$_{\odot}$ has been extracted consistently in many recent studies in the literature. Using representative $R_{1.4}$ data, we infer high-density nuclear symmetry energy $E_{\rm{sym}}(\rho)$ and the associated nucleon specific energy $E_0(\rho)$ in symmetric nuclear matter (SNM) within a Bayesian statistical approach using an explicitly isospin-dependent parametric Equation of State (EOS) for nucleonic matter. We found that:
(1) The available astrophysical data can already improve significantly our current knowledge about the EOS in the density range of $\rho_0-2.5\rho_0$. In particular, the symmetry energy at twice the saturation density $\rho_0$ of nuclear matter is determined to be $E_{\mathrm{sym}}(2\rho_0)$ =39.2$_{-8.2}^{+12.1}$ MeV at 68\% confidence level. 
(2) A precise measurement of the $R_{1.4}$ alone with a 4\% 1$\sigma$ statistical error but no systematic error will not improve much the constraints on the EOS of dense neutron-rich nucleonic matter compared to what we extracted from using the available radius data. (3) The $R_{1.4}$ radius data and other general conditions, such as the observed NS maximum mass and causality condition introduce strong correlations for the high-order EOS parameters. Consequently, the high-density behavior of $E_{\rm{sym}}(\rho)$ inferred depends strongly on how the high-density SNM EOS $E_0(\rho)$ is parameterized, and vice versa. (4) The value of the observed maximum NS mass and whether it is used as a sharp cut-off for the minimum maximum mass or through a Gaussian distribution affect significantly the lower boundaries of both the $E_0(\rho)$ and $E_{\rm{sym}}(\rho)$ only at densities higher than about $2.5\rho_0$. 
\end{abstract}
\maketitle
\newpage

\section{Introduction}\label{sec1}
The energy per nucleon $E(\rho ,\delta )$ (also referred as nucleon specific energy) in nuclear matter at nucleon density $\rho$, isospin asymmetry $\delta\equiv (\rho_n-\rho_p)/\rho$ and zero temperature is the most basic input for calculating the Equation of State (EOS) of cold neutron star (NS) matter regardless of the complexity of the models used. It can be well approximated by the isospin-parabolic expansion \citep{Bom91}
\begin{equation}\label{eos0}
E(\rho ,\delta )=E_0(\rho)+E_{\rm{sym}}(\rho )\cdot \delta ^{2} +\mathcal{O}(\delta^4),
\end{equation}
where $E_0(\rho)$ is the energy per nucleon in symmetric nuclear matter (SNM) having equal numbers of neutrons and protons while the symmetry energy $E_{\rm{sym}}(\rho)$  encodes the energy associated with the neutron-richness of the system. On one hand, much progress has been made over the last few decades in constraining the SNM EOS $E_0(\rho)$ not only around but also significantly above the saturation density of nuclear matter $\rho_0\approx 2.8\times 10^{14}$ g/cm$^{3}$ (corresponding to $\rho_0\approx 0.16~\textrm{nucleons}/\textrm{fm}^3$) by combining the knowledge gained from analyzing both astrophysical observations and terrestrial nuclear experiments, see, e.g., refs. \citep{Danielewicz02,Oertel17,Herman17,Garg18,ISAAC18,BUR18,Zhang19apj}.  On the other hand, while the symmetry energy $E_{\rm{sym}}(\rho )$ mostly around and below $\rho_0$ has been relatively well constrained in recent years \citep{Tesym}, very little is known about the symmetry energy at supra-saturation densities. In fact, it has been broadly recognized that the high-density nuclear symmetry energy is presently among the most important but undetermined quantities reflecting the still mysterious nature of dense neutron-rich nuclear matter \citep{ditoro,Steiner05,LCK08,Lattimer2012,Tsang12,Chuck14,Bal16,Li17}.  Thus not surprisingly, to determine the density dependence of nuclear symmetry energy was identified as a major scientific thrust for nuclear astrophysics in both the U.S. 2015 Long Range Plan for Nuclear Sciences \citep{LRP2015} and the Nuclear Physics European Collaboration Committee (NuPECC) 2017 Long Range Plan \citep{NuPECC}.

Both the magnitude and slope of nuclear symmetry energy contribute to the pressure of NS matter. For example, the pressure of $npe$ matter in NSs at $\beta$ equilibrium at density $\rho$ and isospin asymmetry $\delta$ is explicitly
\begin{equation}\label{pre-npe}
  P(\rho, \delta)=\rho^2[\frac{dE_0(\rho)}{d\rho}+\frac{dE_{\rm{sym}}(\rho)}{d\rho}\delta^2]+\frac{1}{2}\delta (1-\delta)\rho\cdot E_{\rm{sym}}(\rho).
\end{equation}
The first term is the SNM pressure $P_0(\rho)=\rho^2\frac{dE_0(\rho)}{d\rho}$ while the last two terms are the isospin-asymmetric pressure $P_{\rm{asy}}(\rho,\delta)=\rho^2\frac{dE_{\rm{sym}}(\rho)}{d\rho}\delta^2+\frac{1}{2}\delta (1-\delta)\rho\cdot E_{\rm{sym}}(\rho)$ from nucleons and electrons, separately. At the saturation density $\rho_0$, the $P_0$ vanishes and the electron contribution is also negligible leaving the total pressure determined completely by the slope of the symmetry energy. Both the $P_0$ and $P_{\rm{asy}}$ increase with density with rates determined separately by the respective density dependences of the SNM EOS and the symmetry energy. In the region around $\rho_0\sim 2.5\rho_0$, the $P_{\rm{asy}}$ dominates over the $P_0$ using most EOSs available. At higher densities, the SNM pressure $P_0$ dominates while the
$P_{\rm{asy}}$ also plays an important role depending on the high-density behaviors of nuclear symmetry energy \citep{LiSteiner}.
The exact transition of dominance from $P_{\rm{asy}}$ to $P_0$ depends on the stiffnesses of both the SNM EOS and the symmetry energy. It is also well known that the radius $R_{1.4}$ of canonical NSs is essentially determined by the pressure at densities around $\rho_0\sim 2.5\rho_0$ \citep{Lattimer00} while the maximum mass of NSs is determined by the pressure at higher densities reached in the core. Thus, the knowledge about the density dependence of nuclear symmetry energy is important for understanding measurements of both the masses and especially the radii of NSs. Moreover, the critical densities for forming hyperons \citep{Toki,Lee,Kubis1,Pro19},  $\Delta(1232)$ resonances \citep{Italy,Cai-D,Ang,India,Ramos}, kaon condensation \citep{Kub} and the quark phase \citep{Ditoro2,Wu19} are also known to depend sensitively on the high-density nuclear symmetry energy. Information about the latter is thus a prerequisite for exploring the evolution of NS matter phase diagram in the isospin dimension. Once the $E_{\rm{sym}}(\rho)$ is better determined and hopefully with more astrophysical data, it would be interesting to introduce extra model parameters characterizing the physics associated with the exotic particles and/or new phases predicted to appear in super-dense neutron-rich matter. With the very limited data available and expensive computational costs of simultaneously inferring a lot more than the six EOS parameters we already have in the minimum NS model consisting of only nucleons and two leptons, our goals in this work are conservative and practical. However, inferring new physics parameters associated with the exotic degrees of freedom and new phases in super-dense neutron-rich matter from astrophysical data by extending the model used in the present work are high on our working agenda.

To constrain the EOS of ultra-dense neutron-rich nuclear matter has been a longstanding goal of several branches of astrophysics and astronomy. It is a major science driver in building several new research facilities, such as various advanced X-ray observatories and earth-based large telescopes as well as gravitational wave detectors on earth and in space. In particular, ongoing and planned observations \citep{Watts16,Ozel16,Watts19,wp1,wp2} with, e.g., Chandra, XMM-Newton, Neutron Star Interior Composition Explorer [NICER), the upgraded LIGO and VIRGO gravitational wave detector, are all aiming at measuring more precisely the mass-radius correlations of NSs to help constrain the EOS of dense neutron-rich nuclear matter. In the near future, Athena: Advanced Telescope for High-ENergy Astrophysics \citep{Athena}, eXTP: enhanced X-ray Timing and Polarimetry \citep{eXTP}, Large Observatory for x-ray Timing (LOFT-P) \citep{LOFT} and STROBE-X: Spectroscopic Time-Resolving Observatory for Broadband Energy X-rays \citep{Strobe} will further improve precisions of the mass and/or radius measurements. On the other hand, terrestrial nuclear reactions induced by high-energy radioactive ion beams at several new facilities under construction in several countries are also expected to provide improved constraints on the EOS, especially its symmetry energy term, of dense neutron-rich nuclear matter, see, e.g., refs. \citep{FRIB,Hong14,Tamii14,Li17,Wolfgang}.

\begin{deluxetable}{lll}
\tablecolumns{3}
\tablecaption{The radius $R_{1.4}$ data used in this work.}
\tablewidth{0pt}
\tablehead{
\colhead{Radius $R_{1.4}$ (km) (90\% confidence level)} & \colhead{ Source} & \colhead{Reference}
}
\startdata
11.9$^{+1.4}_{-1.4}$&GW170817 & \citep{LIGO18} \\
10.8$^{+2.1}_{-1.6}$ &GW170817 & \citep{De18}  \\
11.7$^{+1.1}_{-1.1}$ &QLMXBs  &\citep{Lattimer14} \\
\hline
$11.9\pm 0.8, 10.8\pm 0.8, 11.7\pm 0.8$ & Imagined case-1 & this work\\
$11.9\pm 0.8$ & Imagined case-2 & this work\\
\enddata
\label{tab:data}
\end{deluxetable}
\begin{figure}[ht]
  \vspace{-2cm}
\begin{center}
  \includegraphics[width=10cm,height=12cm]{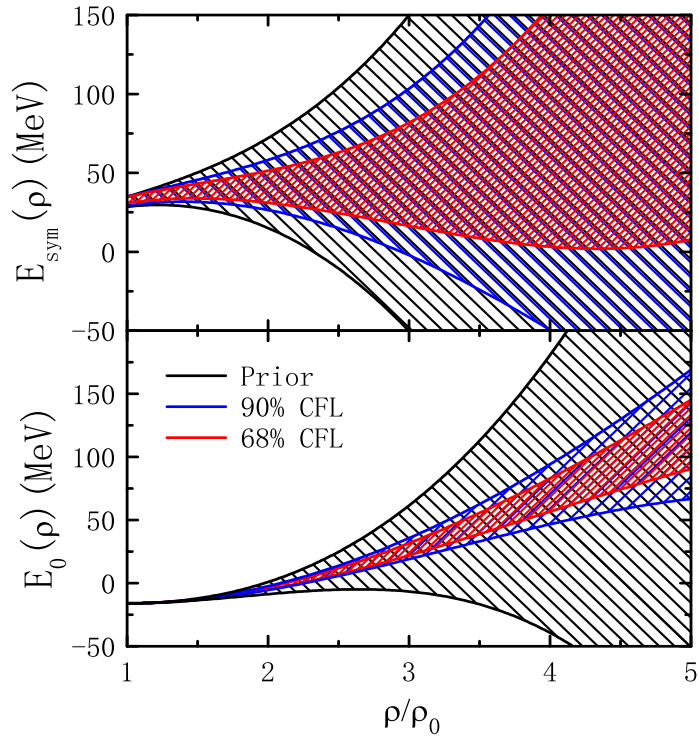}
  \vspace{-1cm}
  \caption{Nuclear symmetry energy (upper) and the energy per nucleon in symmetric nuclear matter (lower) as functions of the reduced density $\rho/\rho_0$. The black, blue and red shadows represent the prior as well as the 90\% and 68\% credible areas of the posterior functions inferred from the Bayesian analysis in this work using the three radius data listed in Table \ref{tab:data} and all known astrophysical constraints about the EOS of neutron star matter, respectively.}\label{E0Esym}
\end{center}
\end{figure}

While waiting eagerly with much interest for results from the new observations and experiments, here we report results of a Bayesian inference of high-density nuclear symmetry energy using the latest $R_{1.4}$ radius data for canonical NSs extracted from analyzing the GW170817 and quiescent low-mass X-ray binaries (QLMXBs) as listed in Table \ref{tab:data} under the condition that any EOS has to be stiff enough to support the observed maximum NS mass, causal and thermodynamically stable at all densities. While there are still some interesting issues and model dependences in extracting the $R_{1.4}$ from the raw data of both gravitational waves and X-rays, within the quoted errors bars  in Table \ref{tab:data} the results from different analyses using various models are largely consistent as discussed in detail in Section 5 of ref. \citep{BALI19}. It is worth emphasizing that while the two NSs involved in GW170817 have significantly different mass ranges, the two independent analyses in refs \citep{LIGO18,De18} using different approaches all found consistently that the two NSs have essentially identical radii within the specified uncertainties in Table \ref{tab:data}. In particular, without any prior restriction on the radii,  \cite{LIGO18} found that the two NSs have the same radius independent of their masses although the radius has different values in different models they used. While in the analyses by \cite{De18}, the two NSs are assumed to have the same radius and they found that the radius is basically independent of the prior mass distributions they assumed for the two NSs. In this study, we took the radii they extracted as for the $R_{1.4}$ consistent with their finding that the radius is mass independent. However, we notice that the error bars of the inferred radii are still large. The possible small mass dependences of the radii predicted by some model EOSs are probably buried in the current uncertainty ranges of the radii.

For comparisons and see how reduced error bars of $R_{1.4}$ data may help us reduce uncertainties of the inferred EOS parameters, we will also use the two imaged cases listed at the bottom of Table \ref{tab:data}. The three real data points have 6-13\% $1\sigma$ statistical errors and their mean values vary by about 1 km from each other, representing approximately a 10\% systematic error of the real case. The imaginary case-1 has the same $R_{1.4}$ mean values as the real data set but a constant absolute error of 0.8 km at 90\% confidence level (CFL). It represents a case of about 4-5\% statistical error of each data point at 68\% CFL but the same systematic error as the real case. While the imaginary case-2 represents a situation where all measurements give the same result as the first data point of imaginary case-1, i.e., no systematic error but a 4\% $1\sigma$ statistical error around the mean value of $R_{1.4}=11.9$ km. Comparisons of results from these 3 cases serve well our main purposes in this work. We would like to establish a generally useful reference for measuring the progress in determining the high-density nuclear symmetry to be brought to us hopefully soon by the new astrophysical observations and terrestrial experiments. Besides being useful for screening predictions of nuclear many-body theories, the obtained posterior probability distribution functions (PDFs) of six parameters involved in an explicitly isospin-dependent parametric EOS are used to construct the confidence boundaries of both the SNM EOS $E_0(\rho)$ and symmetry energy $E_{\rm{sym}}(\rho)$.

For busy readers who are eager to know quickly the most important conclusions of our work, shown in Figure \ref{E0Esym} \footnote{The $E_0(\rho)$ and $E_{\rm{sym}}(\rho)$ shown here should be used jointedly in Eq. \ref{eos0} with the $\delta$ determined self-consistently by the $\beta$ equilibrium and charge neutrality conditions. As we shall discuss, when the $E_{\rm{sym}}(\rho)$ is zero or negative, the system becomes pure neutron matter. The negative value along the lower boundary of $E_0(\rho)$ at high densities does not necessarily mean that SNM at these densities are stable and energetically more preferred.} is the inferred symmetry energy $E_{\rm{sym}}(\rho)$ and the SNM EOS $E_0(\rho)$ as functions of the reduced density $\rho/\rho_0$ using the real data set. The black, blue and red shadows represent the prior, 90\% and 68\% credible areas, respectively. Compared to what we currently know (the prior functions) mostly based on nuclear experiments and theories, the refinement brought in by the astrophysical observations is very significant. More specifically, at densities between $\rho_0-2.5\rho_0$,  both the $E_0(\rho)$ and $E_{\rm{sym}}(\rho)$ are constrained reasonably tightly within their respective narrow bands approximately independent of whether we parameterize the $E_0(\rho)$ with 2 or 3 terms and the $E_{\rm{sym}}(\rho)$ with 3 or 4 terms, respectively.
In particular, the symmetry energy at 2$\rho_0$ is determined to be $E_{\mathrm{sym}}(2\rho_0)$ =39.2$_{-8.2}^{+12.1}$ MeV at 68\% CFL.
However, at densities above about $2.5\rho_0$, the 68\% confidence boundaries for both the $E_0(\rho)$ and $E_{\rm{sym}}(\rho)$ diverge depending strongly on the EOS parameterizations used.

The rest of the paper is organized as follows: in the next section we outline the explicitly isospin-dependent parametric EOS for modeling NSs containing neutrons, protons, electrons and muons (the $npe\mu$ matter). We then outline the steps we used within the Markov-Chain Monte Carlo (MCMC) approach to sample the posterior PDFs of six EOS parameters. In section 3, we present and discuss the posterior PDFs and correlations of the EOS parameters under various conditions to explore possible model dependences as well as effects of uncertain factors and data accuracies. We then construct the 68\% confidence boundaries of both the SNM EOS $E_0(\rho)$ and symmetry energy $E_{\rm{sym}}(\rho)$. Finally, we summarize our main findings.

\section{Theoretical framework}
\label{sec2}
\begin{deluxetable}{lcc}
\tablecolumns{3}
\tablecaption{Prior ranges of the six EOS parameters used}
\tablewidth{0pt}
\tablehead{
\colhead{Parameters} & \colhead{Lower limit} & \colhead{Upper limit (MeV)}
}
\startdata
$K_0$ & 220 & 260 \\
$J_0$ & -800 & 400 \\
$K_{\mathrm{sym}}$  & -400 & 100 \\
$J_{\mathrm{sym}}$ & -200 & 800 \\
$L$ & 30 & 90 \\
$E_{\mathrm{sym}}(\rho_0)$ & 28.5 & 34.9 \\
\enddata
\label{tab:pri}
\end{deluxetable}
\subsection{Explicitly isospin-dependent parametric EOS for the cores of neutron stars}
\label{NS-model}
In Bayesian inferences of nuclear EOSs from astrophysical data, various parametric and non-parametric representations of the EOS at supra-saturation densities have been used in the literature, see, e.g. refs. \citep{Steiner10,Alv16,Ozel16b,Raithel17,Riley18,Ra18,Lim2019,Miller19,Landry,Gre19}. While there is relatively little disagreement about the EOSs of the low-density crust and nuclear matter near $\rho_0$. For a very recent review of different ways of parameterizing the EOSs, their advantages, drawbacks, technical applicabilities and proposals of reducing the prior-dependence in Bayesian inference of EOS parameters, we refer the reader to Section 2.4 of ref. \citep{Baiotti} and references therein.
Because the pressure in NSs at $\beta$ equilibrium is a function only of the density, namely the pressure $P(\rho)$ is barotropic, most studies parameterize the pressure directly as a function of density using piecewise-polytropic parameterizations \citep{Lee00}, the spectral parameterizations \citep{Lee18} or parameterizations generated with a Gaussian process \citep{Landry}. Indeed, the EOS given in terms of $P(\rho)$ is enough for solving the Tolman-Oppenheimer-Volkov (TOV) equations \citep{Tolman34,Oppenheimer39}
to obtain a mass-radius sequence. However, to obtain accurate information about the high-density nuclear symmetry energy and the corresponding density profile of proton fraction $x_p(\rho)$ in the cores of NSs at $\beta$ equilibrium, one has to construct the pressure $P(\rho)$ by parameterizing directly the underlying $E_0(\rho)$ and $E_{\rm{sym}}(\rho)$, separately. Since the $x_p(\rho)$ in NSs at $\beta$ equilibrium is uniquely determined by the $E_{\rm{sym}}(\rho)$ through the chemical equilibrium and charge neutrality conditions, the pressure $P(\rho)$ can be easily constructed from the parameterized  $E_0(\rho)$ and $E_{\rm{sym}}(\rho)$. Such procedure has been used in a number of studies for several purposes in the literature, see, e.g., refs. \citep{Oyamatsu07,Sotani12,MM1,MM2,Zhang18,Malik18,Lim2019,Armen19}, albeit sometimes using different numbers of parameters for either one or both of the $E_0(\rho)$ and $E_{\rm{sym}}(\rho)$. For a recent review, see, ref. \citep{BALI19}.
In this work, we use this procedure in preparing the NS EOS for our Bayesian inference of the high-density nuclear symmetry energy. Details of constructing this NS EOS model were given in refs. \citep{Zhang18,Zhang19a,Zhang19jpg,Zhang19apj}.
In the previous work, however, we fixed the low-order EOS parameters at their currently known most probable values, then several NS observables were inverted in the three-dimensional high-density EOS parameter space. In the Bayesian analyses here, we infer the PDFs of all six EOS parameters from the $R_{1.4}$ data discussed in the previous section. Thus, in the remainder of this section, we only summarize the parts of our EOS parameterization most relevant for the Bayesian analyses.

Within the minimal model of NSs consisting of neutrons, protons, electrons and muons at $\beta$ equilibrium, the pressure
\begin{equation}\label{pressure}
  P(\rho, \delta)=\rho^2\frac{d\epsilon(\rho,\delta)/\rho}{d\rho}
\end{equation}
is determined by the energy density
\begin{equation}\label{lepton-density}
  \epsilon(\rho, \delta)=\rho [E(\rho,\delta)+M_N]+\epsilon_l(\rho, \delta),
\end{equation}
where $M_N$ represents the average nucleon mass and $\epsilon_l(\rho, \delta)$ denotes the lepton energy density. As shown by Eq. (\ref{eos0}), the nucleon energy $E(\rho,\delta)$ is given by the SNM EOS $E_0(\rho)$ and the symmetry energy $E_{\rm{sym}}(\rho)$ which are parameterized respectively according to
\begin{equation}\label{E0-taylor}
  E_{0}(\rho)=E_0(\rho_0)+\frac{K_0}{2}(\frac{\rho-\rho_0}{3\rho_0})^2+\frac{J_0}{6}(\frac{\rho-\rho_0}{3\rho_0})^3,
\end{equation}
where $E_0(\rho_0)$=-15.9 MeV at $\rho_0=0.16~\textrm{nucleons}/\textrm{fm}^3$ (both the $E_0(\rho_0)$ and $\rho_0$ are fixed at these values in this work) and
\begin{equation}\label{Esym-taylor}
    E_{\rm{sym}}(\rho)=E_{\rm{sym}}(\rho_0)+L(\frac{\rho-\rho_0}{3\rho_0})+\frac{K_{\rm{sym}}}{2}(\frac{\rho-\rho_0}{3\rho_0})^2\nonumber
  +\frac{J_{\rm{sym}}}{6}(\frac{\rho-\rho_0}{3\rho_0})^3.
\end{equation}
As discussed earlier \citep{Zhang18,Zhang19a}, the above parameterizations naturally become the Taylor expansions in the limit of $\rho\rightarrow \rho_0$. Compared to the widely used multi-segment polytropic EOSs often with connecting densities/pressures of different density regions as parameters, the asymptotic boundary conditions of the above parameterizations near $\rho_0$ and $\delta=0$ facilitate the use of prior knowledge on the EOS provided by terrestrial nuclear laboratory experiments and/or nuclear theories. Near $\rho_0$ when the above parameterizations become Taylor expansions of the nuclear energy density functionals, the EOS parameters start obtaining their asymptotic physical meanings. Namely, the $K_0$ parameter becomes the incompressibility of SNM $K_0=9\rho_0^2[\partial^2 E_0(\rho)/\partial\rho^2]|_{\rho=\rho_0}$ and the $J_0$ parameter becomes the skewness of SNM $J_0=27\rho_0^3[\partial^3 E_0(\rho)/\partial\rho^3]|_{\rho=\rho_0}$ at saturation density, while the four parameters involved in the $E_{\rm{sym}}(\rho)$ become the magnitude $E_{\rm{sym}}(\rho_0)$, slope $L=3\rho_0[\partial E_{\rm{sym}}(\rho)/\partial\rho]|_{\rho=\rho_0}$, curvature  $K_{\rm{sym}}=9\rho_0^2[\partial^2 E_{\rm{sym}}(\rho)/\partial\rho^2]|_{\rho=\rho_0}$ and skewness $J_{\rm{sym}}=27\rho_0^3[\partial^3 E_{\rm{sym}}(\rho)/\partial\rho^3]|_{\rho=\rho_0}$ of nuclear symmetry energy at saturation density, respectively. These connections between the EOS parameters and the characteristics of nuclear matter and symmetry energy at $\rho_0$ provide us naturally some useful information about the EOS parameters. Summarized in Table \ref{tab:pri} are the currently known
ranges of the asymptotic values of the EOS parameters near $\rho_0$ based on systematics of terrestrial nuclear experiments and predictions of various nuclear theories. As indicated in
Table \ref{tab:pri}, while the $K_0$, $E_{\mathrm{sym}}(\rho_0)$ and $L$ have been constrained to relatively small ranges \citep{Danielewicz02,Li13,Oertel17,Shlomo06,Piekarewicz10},  the $J_0$, $K_{\mathrm{sym}}$ and $J_{\mathrm{sym}}$ describing the EOS of dense neutron-rich matter are only poorly known in a wide range \citep{Tews17,Zhang17}. In the Bayesian inferences of their PDFs from astrophysical data, we use these ranges as the prior ranges.
We use uniform prior PDFs within these ranges for the six EOS parameters as there is no known physical preference for any specific values of these parameters within the ranges specified. For example, the ranges of $E_{\mathrm{sym}}(\rho_0)$ and $L$ were obtained from the systematics of 53 different analyses of some terrestrial experiments and astrophysical observations \citep{Li13,Oertel17,BALI19}. At present and to our best knowledge, one can only reasonably assume {\it apirori} that the EOS parameters can equally take any value within the ranges listed.

\begin{figure}[ht]
\begin{center}
  \includegraphics[width=5.5cm]{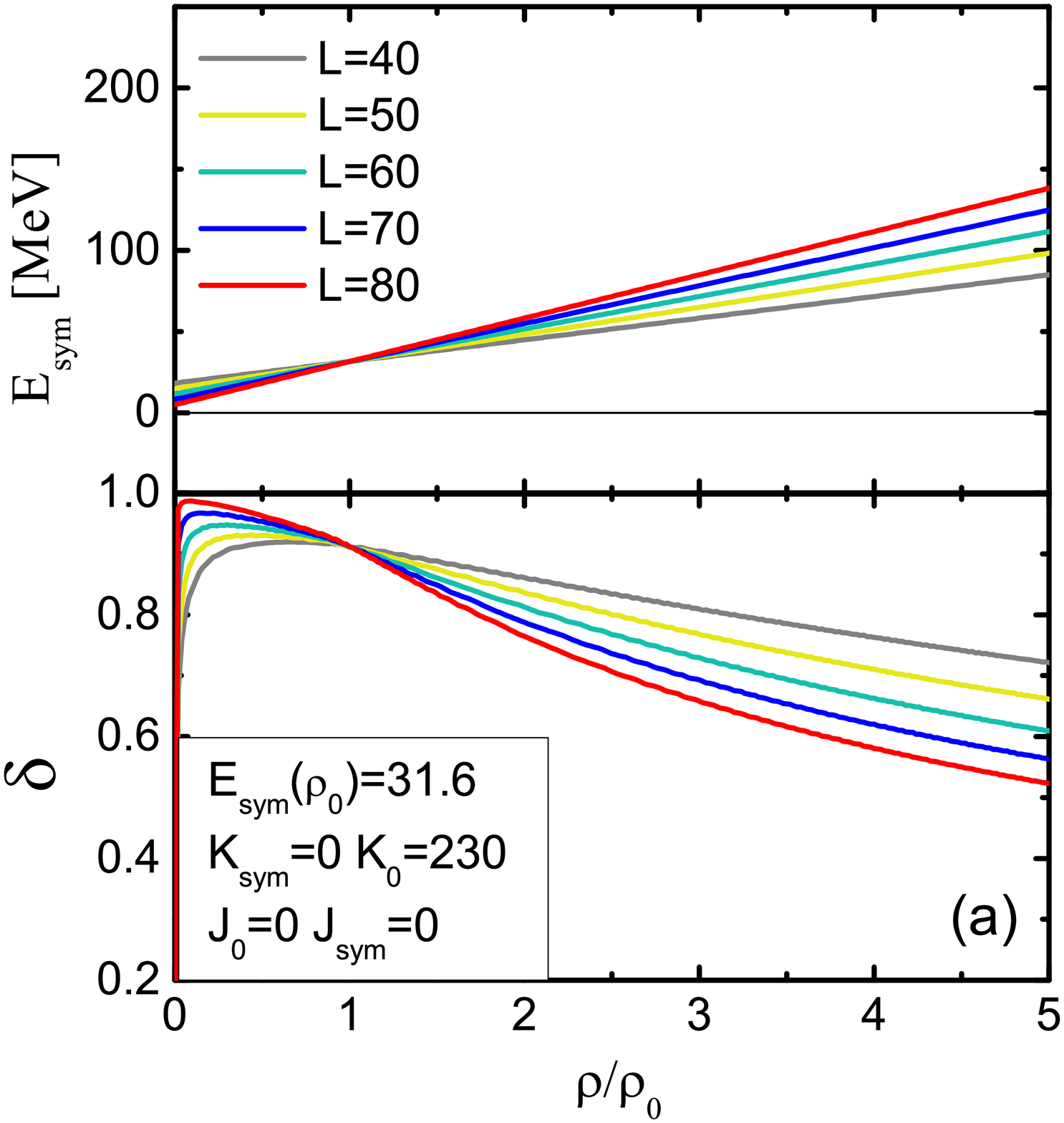}
  \includegraphics[width=5.5cm]{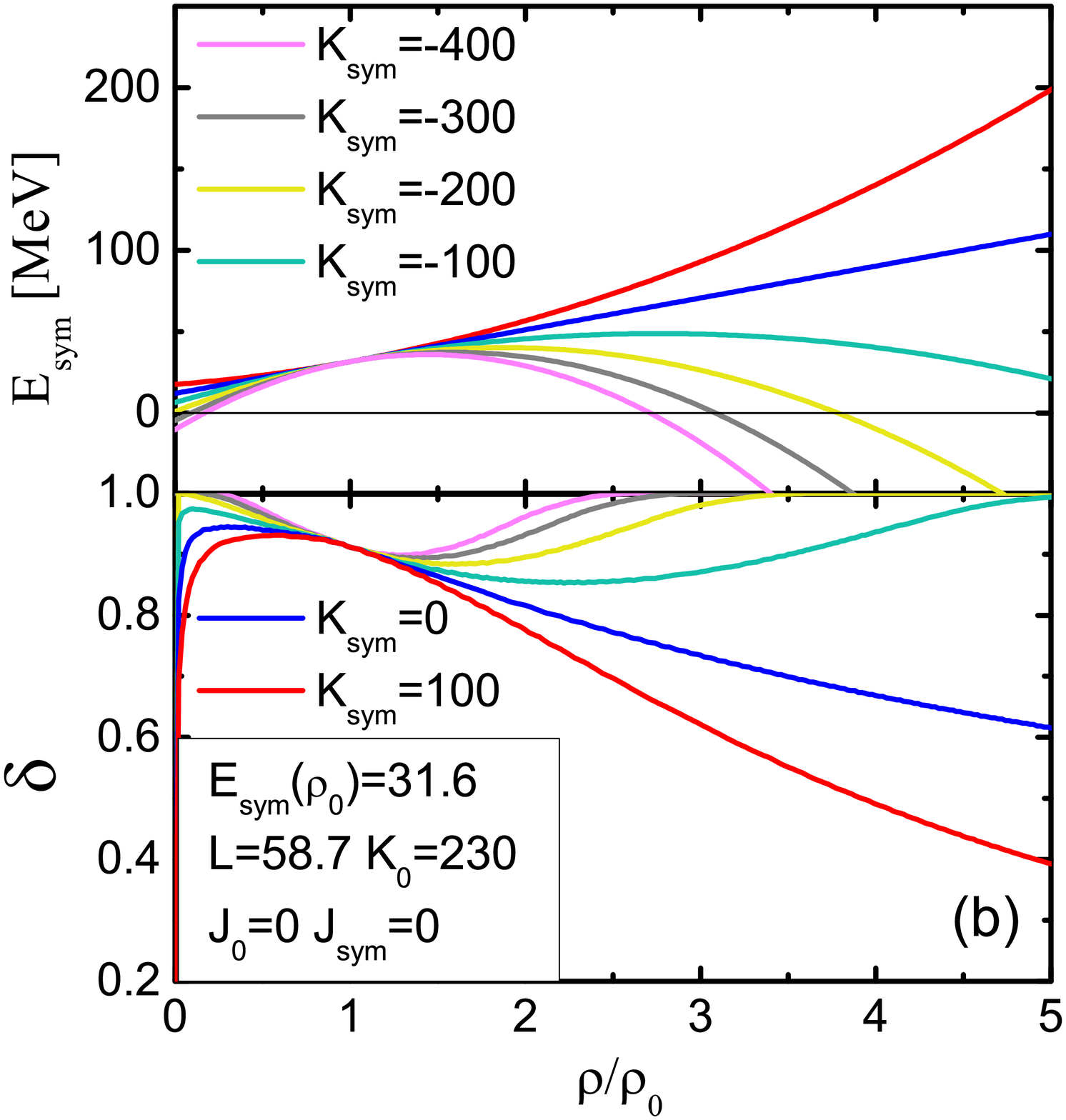}
  \includegraphics[width=5.5cm]{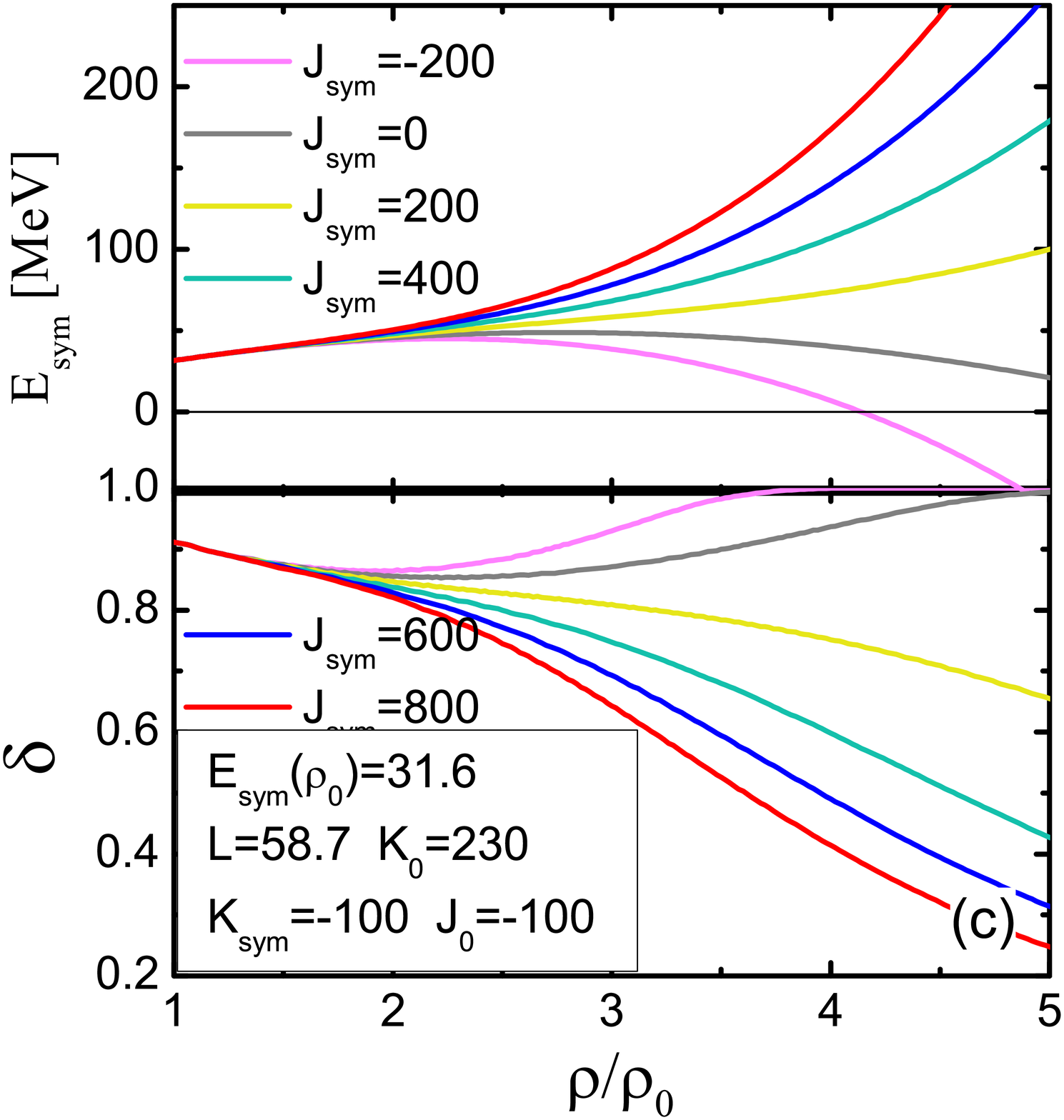}
  \caption{The symmetry energy $E_{\rm{sym}}(\rho)$ and isospin asymmetry $\delta(\rho)$ in neutron star matter at $\beta$-equilibrium as a function of the reduced density $\rho/\rho_0$ for $L=40$, 50, 60, 70, and 80 MeV (a), $K_{\rm{sym}}=-400$, -300, -200, -100, 0, and 100 MeV (b), and $J_{\rm{sym}}=-200$, 0, 200, 400, 600 and 800 MeV (c), respectively. Taken from ref. \citep{Zhang18}.}\label{xfraction}
\end{center}
\end{figure}

The two parameterizations of Eq. (\ref{E0-taylor}) and Eq. (\ref{Esym-taylor}) are combined using Eq. (\ref{eos0}) to obtain the pressure $P(\rho)$ as a function of density only once the density-profile of the isospin asymmetry $\delta(\rho)$ (or the proton fraction $x_p(\rho)$) is calculated self-consistently from the $\beta$-equilibrium condition
$
  \mu_n-\mu_p=\mu_e=\mu_\mu\approx4\delta E_{\rm{sym}}(\rho)
$
and the charge neutrality condition
$  \rho_p=\rho_e+\rho_\mu.
$
The chemical potential for a particle $i$ is obtained from
$
  \mu_i=\partial\epsilon(\rho,\delta)/\partial\rho_i.
$
As an illustration, shown in Fig.\ \ref{xfraction} are samples of the $E_{\rm{sym}}(\rho)$ and the corresponding $\delta(\rho)$ values generated by varying individually the $L$, $K_{\rm{sym}}$ and $J_{\rm{sym}}$ parameters while all other parameters are fixed as indicated in the figure:  (a) $L=40$, 50, 60, 70, and 80 MeV, (b) $K_{\rm{sym}}=-400$, -300, -200, -100, 0, and 100 MeV, and (c) $J_{\rm{sym}}=-200$, 0, 200, 400, 600 and 800 MeV. Clearly, very diverse behaviors of the $E_{\rm{sym}}(\rho)$ and the corresponding $\delta(\rho)$ can be sampled. As expected, the $L$, $K_{\rm{sym}}$ and $J_{\rm{sym}}$ affect gradually more significantly the high-density behavior of the symmetry energy.
Interestingly, because of the $E_{\rm{sym}}(\rho)\cdot \delta^2$ term in the nuclear energy density functional, a higher value of $E_{\rm{sym}}(\rho)$ will lead to a smaller $\delta(\rho)$ at $\beta$ equilibrium. The ramifications of this effect, regarded as the isospin fractionation, is well known in nuclear physics and has been studied extensively in heavy-ion collisions at intermediate energies, see, e.g., refs. \citep{Muller95,Xu00,LCK08}.

Theoretically, the parameterization of $E_{\rm{sym}}(\rho)$ may not always approach zero at zero density when all of its parameters are randomly generated.  While one can cure this completely by introducing additional parameters or use different forms at very low densities, see, e.g., refs. \citep{MM1,Holt18}, practically we avoided this problem by adopting the NV EOS \citep{Negele73} for the inner crust and the BPS EOS \citep{Baym71b} for the outer crust, respectively.  The crust-core transition density and pressure are determined by investigating the thermodynamical instability of the uniform matter in the core \citep{Zhang18}. When the incompressibility of $npe\mu$ matter \citep{Kubis04,Kubis07,Lattimer07}
\begin{equation}\label{tPA}
K_\mu=\rho^2\frac{d^2E_0}{d\rho^2}+2\rho\frac{dE_0}{d\rho}+\delta^2\left[\rho^2\frac{d^2E_{\rm{sym}}}{d\rho^2}+2\rho\frac{dE_{\rm{sym}}}{d\rho}-2E^{-1}_{\rm{sym}}(\rho\frac{dE_{\rm{sym}}}{d\rho})^2\right]
\end{equation}
becomes negative at low densities, the uniform matter becomes unstable against the formation of clusters. Numerical examples of the crust-core transition density and pressure
can be found in ref. \citep{Zhang18}. Clearly, the $K_\mu$ depends mainly on the $K_0$, $L$ and $K_{\rm{sym}}$. For fast MCMC samplings, we have prepared a table of the
crust-core transition density and pressure as functions of $K_0$, $L$ and $K_{\rm{sym}}$ with a bin size of 5 MeV. The latter defines the internal energy resolution of
our Bayesian inference. This table is available up on request from the authors of this work.

\subsection{Bayesian Inference Approach}\label{Bayes}
As discussed above and also in refs. \citep{Zhang18,Zhang19a}, the Eqs. (\ref{E0-taylor}) and (\ref{Esym-taylor}) have the dual meanings (either as parameterizations or Taylor expansions) near the saturation density, but when applied to high densities, they are simply parameterizations. In refs. \citep{Zhang18,Zhang19a}, by fixing the three low-order parameters $K_0$, $E_{\rm{sym}}(\rho_0)$ and $L$ at their most probable values currently known, effects of only the three high-density (order) parameters $J_0$, $K_{\rm{sym}}$ and $J_{\rm{sym}}$ were studied by inverting individually several observables directly in the three-dimensional high-density EOS parameter space. While the results are very useful, not only effects of uncertainties of the low-order parameters but also correlations among the EOS parameters were not considered or treated on equal footing. In the present work, we treat the six parameters in Eqs. (\ref{E0-taylor}) and (\ref{Esym-taylor}) all as free parameters to be constrained simultaneously by the same astrophysical data and several known constraints using the Bayesian inference approach.

The key in this approach is the calculation of the posterior PDFs of the model EOS parameters through the MCMC sampling. For completeness, we recall here the Bayes' theorem
\begin{equation}\label{Bay1}
P({\cal M}|D) = \frac{P(D|{\cal M}) P({\cal M})}{\int P(D|{\cal M}) P({\cal M})d\cal M},
\end{equation}
where $P({\cal M}|D)$ is the posterior probability for the model $\cal M$ given the data set $D$, which is what we are seeking here, while $P(D|{\cal M})$ is the likelihood function or the conditional probability for a given theoretical model $\cal M$ to predict correctly the data $D$, and $P({\cal M})$ denotes the prior probability of the model $\cal M$ before being confronted with the data. The denominator in Eq. (\ref{Bay1}) is the normalization constant.
\begin{figure}[ht]
\begin{center}
  \includegraphics[width=15cm]{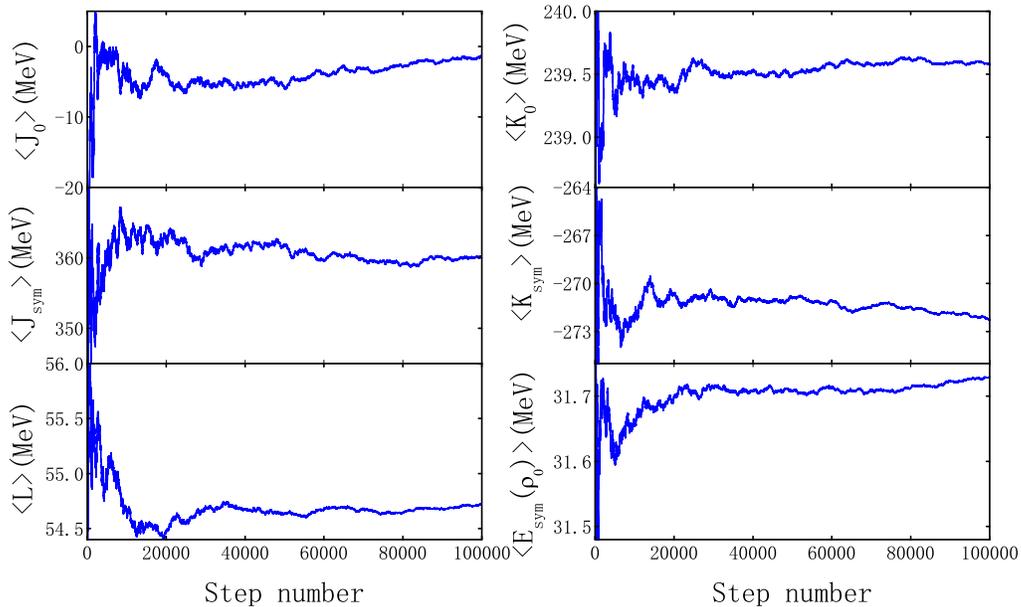}
  \caption{Mean values of the six EOS parameters as functions of the step number.}\label{burn-in}
\end{center}
\end{figure}
We sample uniformly each EOS parameter $p_i$ randomly between its minimum $p_{\mathrm{min},i}$ and maximum $p_{\mathrm{max},i}$ values given in Table \ref{tab:pri}
according to $p_i=p_{\mathrm{min},i}+(p_{\mathrm{max},i}-p_{\mathrm{min},i})x$ with the random number $x$ between 0 and 1. Using the generated EOS parameters, $p_{i=1,2\cdots 6}$, one can construct the NS EOS model ${\cal M}(p_{1,2,\cdots 6})$ as we outlined in the previous subsection. The NS mass-radius sequence is then obtained by solving the TOV equations. The resulting theoretical radius $R_{\mathrm{th},j}$ is subsequently used to calculate the likelihood of this EOS parameter set with respect to the observed radius $R_{\mathrm{obs},j}$ with $j=1,2,3$ in the data set $D(R_{1,2, 3})$ given in Table \ref{tab:data} according to
\begin{equation}\label{Likelihood}
  P[D(R_{1,2,3})|{\cal M}(p_{1,2,\cdots 6})]=\prod_{j=1}^{3}\frac{1}{\sqrt{2\pi}\sigma_{\mathrm{obs},j}}\exp[-\frac{(R_{\mathrm{th},j}-R_{\mathrm{obs},j})^{2}}{2\sigma_{\mathrm{obs},j}^{2}}],
\end{equation}
where $\sigma_{\mathrm{obs},j}$ is the $1\sigma$ error bar of the observation $j$. For the radius data given in Table 1, the upper/lower error bar in radius at 90\% CFL is $1.645\sigma$. For the asymmetric case with different upper and lower error bars, we took their average before calculating the $\sigma$ value. 
In applying the approach to canonical NSs with the same mass of 1.4 M$_{\odot}$, $R_{\mathrm{th},j}\equiv R_{1.4}^{\mathrm{th}}$ is independent of the index $j$, i.e., we treat the three radii in the data set as from independent observations of the same NS.
We will calculate both the marginalized PDFs of all individual EOS parameters and the two-parameter correlations by integrating over all other parameters.
In particular, the PDF of a model parameter $p_i$ is analytically given by
\begin{equation}\label{Bay3}
P(p_i|D) = \frac{\int P(D|{\cal M}) dp_1dp_2\cdots dp_{i-1}dp_{i+1}\cdots dp_6}{\int P(D|{\cal M}) P({\cal M})dp_1dp_2\cdots dp_6},
\end{equation}
while the correlation function between any two parameters $p_i$ and $p_j$ can be similarly written as an integral of the $P(D|{\cal M})$ over all parameters except the 
$p_i$ and $p_j$ themselves. Numerically,  the $P[D(R_{1,2,3})|{\cal M}(p_{1,2,\cdots 6})]$ is simulated using the MCMC sampling approach. The integrations are also done naturally using the Monte Carlo approach. Namely, the PDF of a model parameter $p_i$ is obtained by summing up the accepted $P(D|{\cal M})$ at all steps in the MCMC process and the sum is binned only according to the value of $p_i$ regardless of all other parameters. While the two-dimensional correlation function between the two parameters $p_i$ and $p_j$ is obtained by binning the sum according to both the $p_i$ and $p_j$ regardless of the values of all other parameters. While it is necessary to normalize the PDFs of single EOS parameters individually, unnormalized correlation functions are sufficient for all practical purposes.

Many MCMC techniques are available in the literature, in this work we adopted the Metropolis-Hastings algorithm \citep{Metropolis53,Hastings70}. In developing our MCMC package in Fortran 90, we used as an example the MCMC code developed in ref. \citep{Kim09} following the procedure given in ref. \citep{Zhang06}. It is well known that because the MCMC process does not start from an equilibrium distribution, initial samples in the so-called burn-in period have to be thrown away \citep{Trotta17}. The adequate burn-in step numbers are those after which either the posterior densities or means of the model parameters remain approximately constants on their trace plots during the MCMC sampling. Shown in Figure \ref{burn-in} are the mean values for our six EOS parameters as functions of the step number. It is seen that after about 40,000 steps, the mean values for all of the parameters stay approximately constants. Thus, we take 40,000 burn-in steps in all calculations performed in this work.

\begin{figure}[ht]
\begin{center}
  \includegraphics[width=13cm]{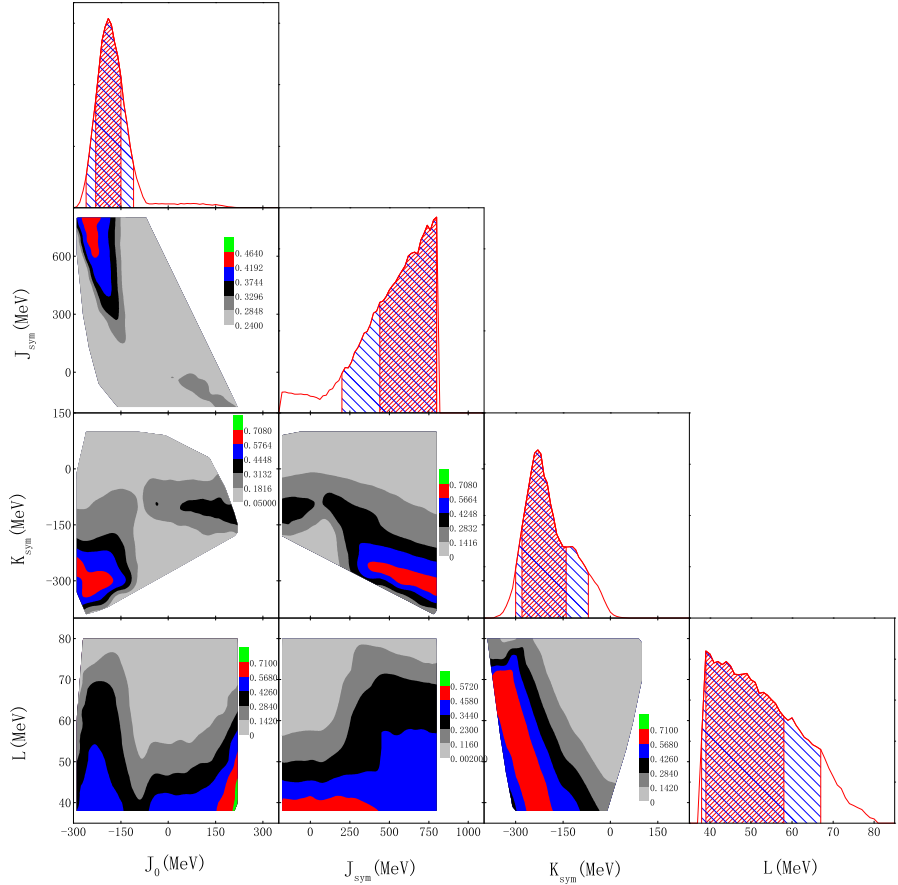}
  \caption{Posterior probability distribution functions and correlations of high-density EOS parameters.}\label{pdf-cor1}
\end{center}
\end{figure}
\begin{figure}[ht]
\begin{center}
  \includegraphics[width=13cm,height=19cm]{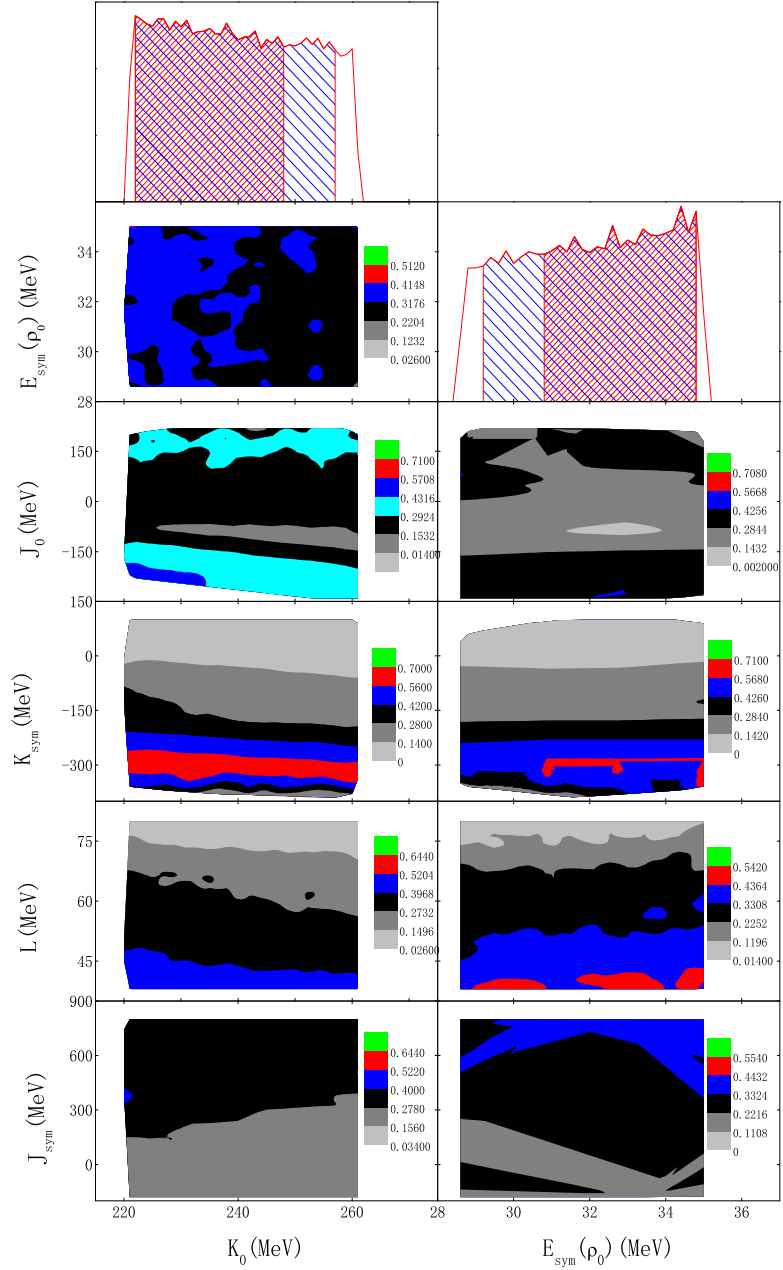}
  \caption{Posterior probability distribution functions of low-density EOS parameters and their correlations with the high-density EOS parameters.}\label{pdf-cor2}
\end{center}
\end{figure}
\begin{table}[tbp]
\centering
\caption{{\protect\small Most probable values of the EOS parameters and their 68\%, 90\% confidence boundaries }}
\label{SixPara}%
\begin{tabular}{ccccccc}
\hline\hline
Parameter (MeV) &$68\%$ boundaries &$90\%$ boundaries\\
\hline
$J_0$ & $-190_{-40}^{+40}$ & $-190_{-70}^{+80}$\\
$K_0$ & $222_{-0}^{+26}$ & $222_{+35}^{-0}$ \\
$J_{\mathrm{sym}}$ & $800_{-360}^{+0}$ & $800_{-600}^{+0}$  \\
$K_{\mathrm{sym}}$ & $-230_{-50}^{+90}$ & $-230_{-70}^{+160}$\\
$L$ & $39_{-0}^{+19}$ & $39_{-1}^{+28}$ \\
$E_{\mathrm{sym}}(\rho_0)$ & $34_{-4.8}^{+0.8}$ & $34_{-3.2}^{+0.8}$  \\ \hline\hline
\end{tabular}%
\end{table}

\section{Results and discussions}
\label{sec2.2}
All EOSs we constructed satisfy the following default conditions:
(i) The crust-core transition pressure stays positive; (ii) The thermaldynamical stability condition, $dP/d\varepsilon\geq0$, is satisfied at all densities; (iii) The causality condition is enforced at all densities; (iv) The generated NS EOS should be stiff enough to support the currently observed maximum mass of NSs. We consider the maximum mass as a variable and discuss its effects on what we infer abut the EOS parameters by using either a sharp cut-off at 1.97, 2.01 or 2.17 M$_{\odot}$ or a Gaussian distribution for the maximum mass centered around 2.01 or 2.17 M$_{\odot}$ with a $1\sigma$ error of 0.04 or 0.11 M$_{\odot}$, respectively. In presenting and discussing our results in the following, we use as a baseline the default result obtained by using the sharp cut-off at 1.97 M$_{\odot}$, namely all EOSs have to be able to support NSs at least as massive as 1.97 M$_{\odot}$, and the three real $R_{1.4}$ radius data given in Table 1. Based on all the information we have so far,  conclusions based on the default results are the most realistic and conservative. Results from variations of the physics conditions and/or using the imagined data sets will then be compared with the default results. We focus on the marginalized PDFs of the model parameters and their correlations induced by the physics conditions and/or the radius data. The $E_0(\rho)$ and $E_{\rm{sym}}(\rho)$ at various confidence levels in relevant cases will be constructed. Since the PDFs of EOS parameters are asymmetric in most cases, in presenting our results we use the highest posterior density interval \citep{Turkkan14} to locate the 68\% (90\%) credible intervals according to
\begin{equation}\label{HPD}
  \int_{p_{i\mathrm{L}}}^{p_{i\mathrm{U}}}\mathrm{PDF}(p_{i})dp_{i}=0.68~ (0.90),
\end{equation}
where ($p_{i\mathrm{L}}$, $p_{i\mathrm{U}}$) is the (lower, upper) limit of the corresponding narrowest interval of the parameter $p_i$ surrounding the maximum value of the $\mathrm{PDF}(p_i)$.

\subsection{The posterior PDFs and correlations of EOS parameters}
Shown in Figures \ref{pdf-cor1} and \ref{pdf-cor2} are the marginalized posterior PDFs of the six EOS parameters and their correlations. The blue and red shadows outline the 90\% and 68\% credible intervals of the PDFs, respectively.  The most probable values of the six EOS parameters and their 68\%, 90\% confidence boundaries are summarized in Table \ref{SixPara}. Several interesting observations can be made:
\begin{itemize}
\item
The $J_0$ and $K_{\mathrm{sym}}$ parameters are relatively well constrained, and larger values of $J_{\mathrm{sym}}$ but smaller values of $L$ are preferred while the saturation parameters $K_0$ and $E_{\mathrm{sym}}(\rho_0)$, as shown in Figure \ref{pdf-cor2}, are essentially not affected by the radius data and the default physics conditions used in this baseline calculation compared to their prior ranges given in Table \ref{tab:pri}. These features are easily understandable. Since the average density in canonical NSs with 1.4 M$_{\odot}$ is not so high, it is difficult to constrain the high-density symmetry energy parameter $J_{\mathrm{sym}}$ by using the $R_{1.4}$ data. As discussed in refs. \citep{Lattimer00,Lattimer01}, the radius of canonical NSs is most sensitive to the pressure around 2$\rho_0$. One thus expects that the $R_{1.4}$ data is most useful for constraining EOS parameters having largest influences on the pressure around 2$\rho_0$. This appears to be case for the $K_{\mathrm{sym}}$ parameter. It plays a pivotal role in determining the pressure at intermediate densities reached in canonical NSs, and thus being narrowed down to a small range by the $R_{1.4}$ data used in this analysis. The parameter $J_0$ characterizing the high-density behavior of the SNM EOS $E_{0}(\rho)$ is sensitive to the NS maximum mass but not to the radius of canonical NSs \citep{Zhang18,Zhang19a}. It is the condition that all EOSs should be stiff enough to support NSs at least as massive as 1.97 M$_{\odot}$ that narrowed down the $J_0$ parameter range. We shall discuss in more detail the individual roles of $J_0$ and $J_{\mathrm{sym}}$ in subsections \ref{E-J0} and \ref{E-Jsym}, separately, as they are not always considered simultaneously in Bayesian inferences of the nuclear EOS in the literature.

\item
Generally speaking, mathematically one expects that a given parameter is most likely to correlate with its nearest neighbors used in parameterizing a specific function, e.g., $K_{\mathrm{sym}}$ with its left neighbor $L$ and right neighbor $J_{\mathrm{sym}}$ in parameterizing the $E_{\rm{sym}}(\rho)$. All of the six EOS parameters started as independent ones, they become correlated basically by satisfying several energy conservation as well as chemical and mechanical equilibrium conditions when the radii and masses from solving the TOV equations are required to reproduce the radius data within certain ranges under the conditions specified. Two near-by parameters in a given function can easily compensate each other to reproduce the same data under the same condition. It is thus not surprising that there are
stronger anti-correlations between $L$ and $K_{\mathrm{sym}}$ as well as between $J_{\mathrm{sym}}$ and $K_{\mathrm{sym}}$, but only very weak correlations between the saturation-density parameters and the high-density parameters.

\begin{figure}[ht]
\begin{center}
  \includegraphics[width=13cm,height=13cm]{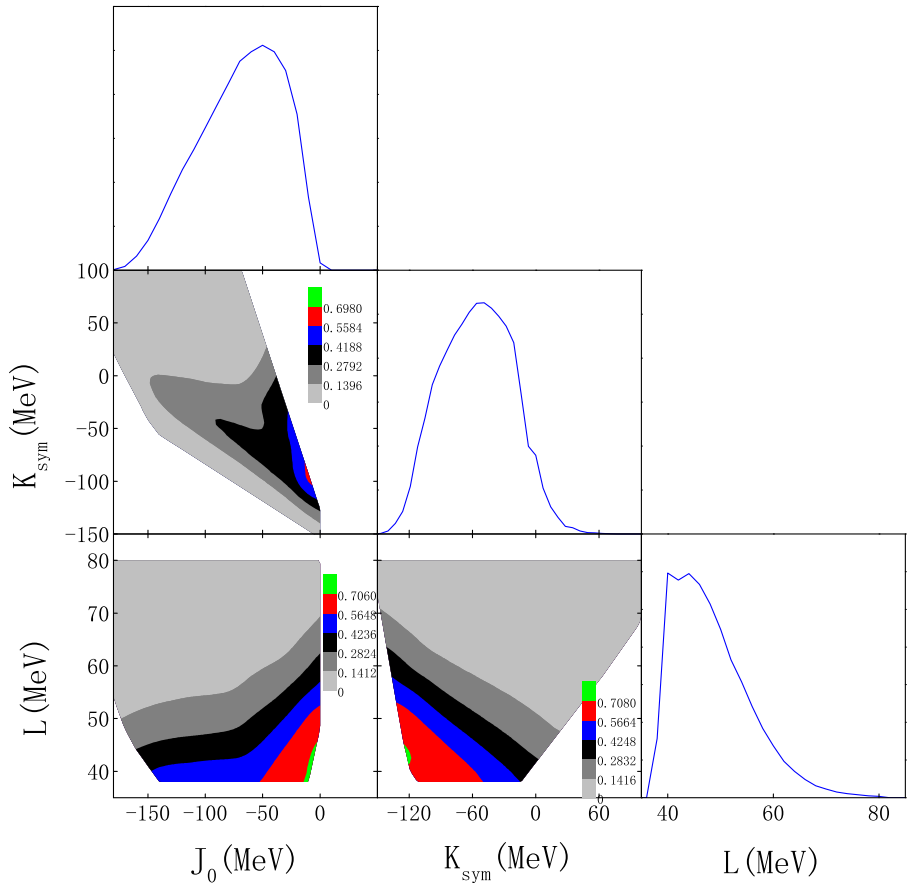}
  \caption{Posterior probability distribution functions and correlations of high-density EOS parameters by setting $J_{\mathrm{sym}}=0$.}\label{Jsym0Cor}
\end{center}
\end{figure}

Two parameters used in parameterizing the $E_0(\rho)$ and $E_{\rm{sym}}(\rho)$ may also be correlated through the total nucleon specific energy $E(\rho,\delta)$ in Eq. (\ref{eos0}). Depending on the resulting value of $\delta$, two parameters from the two parameterizations generally have to be at the same order in density to have strong correlations. For example, the $J_0$ and $J_{\mathrm{sym}}$ both at the third-order in density are expected to be strongly correlated when the $\delta$ is close to 1 achieved when the symmetry energy is super-soft especially with negative $J_{\mathrm{sym}}$ as shown in Fig. \ref{xfraction}. However, correlations between 
low-order (e.g., $K_0$) and high-order (e.g., $J_{\mathrm{sym}}$) parameters are expected to be very weak. In fact, a strong anti-correlation between $J_0$ and $J_{\mathrm{sym}}$ is necessary in order to support the same observed NS maximum mass. Namely, an increasing $J_0$ is needed when $J_{\mathrm{sym}}$ is decreasing to keep the high-density pressure strong enough to support the same massive NSs. In addition, an anti-correlation between $L$ and $K_{\mathrm{sym}}$ exists independent of what values the $J_{\mathrm{sym}}$ takes to keep the same radius $R_{1.4}$.  These features are consistent with those found from studying the constant surfaces of radii, speed of sound and/or the maximum mass in the EOS parameter spaces \citep{Zhang19a,Zhang19jpg}. It is also interesting to note that there is a positive correlation between $J_0$ and $K_{\mathrm{sym}}$. This is understandable from the anti-correlation between $J_0$ and $J_{\mathrm{sym}}$, and the anti-correlation between $J_{\mathrm{sym}}$ and $K_{\mathrm{sym}}$.  However, this finding is different from that found in ref. \citep{Baillot19} where a strong anti-correlation between $J_0$ and $K_{\mathrm{sym}}$ was observed while a similarly strong anti-correlation between $L$ and $K_{\mathrm{sym}}$ was also found as in our present work. We found that this is due to the fact that the $J_{\mathrm{sym}}$ was set to zero in ref. \citep{Baillot19}. In fact, results very similar to theirs are obtained when we set $J_{\mathrm{sym}}$=0 in the present work, as shown in Figure \ref{Jsym0Cor} and to be discussed in more detail in the subsection \ref{E-Jsym}. It is seen that now an anti-correlation between the $J_0$ and $K_{\mathrm{sym}}$ appears while that between the $L$ and $K_{\mathrm{sym}}$ stays the same.

\item Some of the PDFs have extended tails and/or shoulders due to the correlations among the six EOS parameters. For example, the PDF of $J_0$ has an appreciable tail extending to large positive  $J_0$ values. Much efforts have been devoted to predicting the value of $J_0$ over the past years. For example, in ref. \citep{Far97}, $J_0$=-700$\pm$500 MeV was obtained by analyzing nuclear giant monopole resonances. From an earlier Bayesian analysis of some NS data, $J_0$=-280$^{+72}_{-410}$ MeV was inferred by assuming $r_{\mathrm{ph}}\gg R$, where $r_{\mathrm{ph}}$ and $R$ represent the photospheric and stellar radii, respectively, but $J_0$=-500$^{+170}_{-290}$ MeV was obtained with $r_{\mathrm{ph}}= R$ \citep{Steiner10}. By using a correlation analysis method based on the Skyrme Hartree-Fock energy density functional \citep{Lwchen11}, a value of $J_0$=-355$\pm$95 MeV was estimated. While within a nonlinear relativistic mean field model \citep{Cai14}, by combining the constraints imposed by the flow data in heavy-ion collisions \citep{Danielewicz02} and the mass of PSR J0348+0432 \citep{Antoniadis13}, a range of -494 MeV $\leq$ $J_0\leq$ -10 MeV for $J_0$ was inferred.

However,  contrast to the mostly negative values of $J_0$ mentioned above, positive $J_0$ values with very lower probabilities are allowed as indicated in the top left panel of Figure \ref{pdf-cor1}. This is due to the correlations among $J_0$, $J_{\mathrm{sym}}$ and $K_{\mathrm{sym}}$, especially the correlation between $J_0$ and $J_{\mathrm{sym}}$. As discussed in ref. \citep{Zhang18} and illustrated in Figure \ref{xfraction}, extremely small values of $J_{\mathrm{sym}}$ (e.g. negative values of $J_{\mathrm{sym}}$) lead to super-soft symmetry energies. The system then becomes very neutron-rich with $\delta \approx 1$ at high densities. In this case, as shown in Figure \ref{xfraction}, the symmetry energy decreases with increasing density, giving a negative contribution (the second term in Eq. \ref{pre-npe}) to the total pressure. To support the same NSs as massive as 1.97 M$_{\odot}$ or even 1.4 M$_{\odot}$, a much larger contribution to the total pressure from the symmetric part ($E_0(\rho)$ term) is thus required. Because of the weaker correlation between $K_0$ and the data/conditions used, as shown in Figure \ref{pdf-cor2}, it is the $J_0$ that dominates the contribution to the high-density pressure from the $E_0(\rho)$ term. Therefore, the value of $J_0$ has to be positive when $J_{\mathrm{sym}}$ is very small. This conclusion can also be seen in Figure 10 of ref. \citep{Zhang19a} where the constant surface of the maximum mass 2.01 M$_{\odot}$ was examined in the 3-dimensional $J_{\mathrm{0}}$-$K_{\mathrm{sym}}$-$J_{\mathrm{sym}}$ high-density EOS parameter space. We note here that such features were not seen in other Bayesian analyses because the $E_{\rm{sym}}(\rho)$ was not allowed to become super-soft {\it apriori} by setting $J_{\mathrm{sym}}$=0, or often the isospin-independent polytroic EOSs are used at high densities.

It is also interesting to see that there is a small shoulder for the PDF of $K_{\mathrm{sym}}$ around $K_{\mathrm{sym}}\approx$-100 MeV. This is also because of the correlations among $J_0$, $J_{\mathrm{sym}}$ and $K_{\mathrm{sym}}$. As shown in Figure \ref{pdf-cor1}, when $J_0$ has values in the region of 0$\sim$200 MeV, $J_{\mathrm{sym}}$ has negative values while $K_{\mathrm{sym}}$ has values of about -100 MeV with appreciably higher probabilities than the surrounding area on the contour plots of the correlations.

\end{itemize}

\begin{table}[tbp]
\centering
\caption{{\protect\small 68\%, 90\% Credible Intervals and most probable values of the symmetry energy $E_{\rm{sym}}(\rho)$}}
\label{EsymTable}%
\begin{tabular}{ccccccc}
\hline\hline
$\rho/\rho_0$ &$90\%$ Lower Limit &$68\%$ Lower Limit &Most Probable &$68\%$ Upper Limit &$90\%$ Upper Limit \\
\hline
$1.0$ & $29.2$ & $30.8$ & $34.0$ & $34.8$ & $34.8$\\
$1.1$ & $30.3$ & $32.0$ & $35.2$ & $36.7$ & $37.0$ \\
$1.2$ & $31.1$ & $32.8$ & $36.1$ & $38.4$ & $39.2$  \\
$1.3$ & $31.5$ & $33.4$ & $36.9$ & $40.0$ & $41.3$\\
$1.4$ & $31.7$ & $33.7$ & $37.5$ & $41.6$ & $43.4$ \\
$1.5$ & $31.5$ & $33.8$ & $37.9$ & $43.1$ & $45.6$  \\
$1.6$ & $31.1$ & $33.6$ & $38.3$ & $44.7$ & $47.9$  \\
$1.7$ & $30.3$ & $33.2$ & $38.5$ & $46.2$ & $50.2$  \\
$1.8$ & $29.3$ & $32.6$ & $38.8$ & $47.8$ & $52.7$ \\
$1.9$ & $28.0$ & $31.9$ & $38.9$ & $49.5$  & $55.3$  \\
$2.0$ & $26.4$ & $31.0$ & $39.2$ & $51.3$  & $58.2$ \\
$2.1$ & $24.6$ & $29.9$ & $39.4$ & $53.2$  & $61.2$ \\
$2.2$ & $22.5$ & $28.7$ & $39.7$ & $55.3$  & $64.5$  \\
$2.3$ & $20.2$ & $27.4$ & $40.2$ & $57.6$ & $68.1$  \\
$2.4$ & $17.7$ & $26.0$ & $40.7$ & $60.2$ & $72.0$   \\
$2.5$ & $14.9$ & $24.5$ & $41.4$ & $63.0$  & $76.2$  \\
$2.6$ & $11.9$ & $22.9$ & $42.3$ & $66.0$  & $80.8$  \\
$2.7$ & $8.6$ & $21.3$ & $43.4$ & $69.5$ & $85.8$   \\
$2.8$ & $5.2$ & $19.6$ & $44.8$ & $73.2$ & $91.2$   \\
$2.9$ & $1.6$ & $18.0$ & $46.4$ & $77.3$ & $97.1$   \\
$3.0$ & $-2.3$ & $16.3$ & $48.4$ & $81.9$  & $103.4$  \\ \hline\hline
\end{tabular}%
\end{table}

\begin{table}[tbp]
\centering
\caption{{\protect\small 68\%, 90\% Credible Intervals and most probable values for the SNM EOS $E_{0}(\rho)$ }}
\label{E0table}%
\begin{tabular}{ccccccc}
\hline\hline
$\rho/\rho_0$ &$90\%$ Lower Limit &$68\%$ Lower Limit &Most Probable &$68\%$ Upper Limit &$90\%$ Upper Limit\\
\hline
$1.0$ & $-16.0$ & $-16.0$ & $-16.0$ & $-16.0$ & $-16.0$\\
$1.1$ & $-15.9$ & $-15.9$ & $-15.9$ & $-15.9$ & $-15.9$ \\
$1.2$ & $-15.5$ & $-15.5$ & $-15.5$ & $-15.5$ & $-15.4$  \\
$1.3$ & $-14.9$ & $-14.9$ & $-14.9$ & $-14.8$ & $-14.7$\\
$1.4$ & $-14.1$ & $-14.1$ & $-14.1$ & $-13.9$ & $-13.8$ \\
$1.5$ & $-13.1$ & $-13.1$ & $-13.1$ & $-12.7$ & $-12.5$  \\
$1.6$ & $-11.9$ & $-11.9$ & $-11.8$ & $-11.2$ & $-11.0$  \\
$1.7$ & $-10.6$ & $-10.4$ & $-10.4$ & $-9.6$ & $-9.2$  \\
$1.8$ & $-9.0$ & $-8.8$ & $-8.7$ & $-7.7$ & $-7.2$ \\
$1.9$ & $-7.3$ & $-7.0$ & $-6.9$ & $-5.5$ & $-4.9$  \\
$2.0$ & $-5.5$ & $-5.1$ & $-4.8$ & $-3.1$ & $-2.4$  \\
$2.1$ & $-3.5$ & $-3.0$ & $-2.6$ & $-0.6$ & $0.4$ \\
$2.2$ & $-1.3$ & $-0.7$ & $-0.3$ & $2.2$ & $3.4$  \\
$2.3$ & $0.9$ & $1.7$ & $2.3$ & $5.3$ & $6.6$  \\
$2.4$ & $3.3$ & $4.3$ & $5.0$ & $8.5$ & $10.1$   \\
$2.5$ & $5.7$ & $7.0$ & $7.8$ & $11.9$  & $13.8$  \\
$2.6$ & $8.2$ & $9.8$ & $10.8$ & $15.5$  & $17.8$  \\
$2.7$ & $10.8$ & $12.7$ & $13.9$ & $19.3$ & $21.9$   \\
$2.8$ & $13.5$ & $15.7$ & $17.1$ & $23.2$ & $26.3$   \\
$2.9$ & $16.2$ & $18.8$ & $20.5$ & $27.4$ & $30.9$   \\
$3.0$ & $19.0$ & $22.0$ & $24.0$ & $31.7$  & $35.7$  \\
$3.1$ & $21.8$ & $25.2$ & $27.5$ & $36.2$ & $40.7$ \\
$3.2$ & $24.6$ & $28.6$ & $31.2$ & $40.8$ & $45.9$  \\
$3.3$ & $27.5$ & $32.0$ & $35.0$ & $45.6$ & $51.3$  \\
$3.4$ & $30.3$ & $35.4$ & $38.8$ & $50.6$ & $56.9$   \\
$3.5$ & $33.1$ & $38.9$ & $42.8$ & $55.6$  & $62.6$  \\
$3.6$ & $35.9$ & $42.4$ & $46.8$ & $60.9$  & $68.6$  \\
$3.7$ & $38.7$ & $46.0$ & $50.8$ & $66.2$ & $74.7$   \\
$3.8$ & $41.4$ & $49.5$ & $54.9$ & $71.7$ & $81.0$   \\
$3.9$ & $44.1$ & $53.1$ & $59.1$ & $77.3$ & $87.5$   \\
$4.0$ & $46.7$ & $56.7$ & $63.3$ & $83.0$  & $94.2$  \\ \hline\hline
\end{tabular}%
\end{table}

\subsection{The Bayesian inferred high-density EOS confidence intervals}
After obtaining the most probable values and the quantified uncertainties for the EOS parameters, one can derive the corresponding EOS according to Eqs. (\ref{E0-taylor}) and (\ref{Esym-taylor}). Show in Figure \ref{E0Esym} are the symmetry energy $E_{\mathrm{sym}}(\rho)$ and the nucleon specific energy $E_{\mathrm{0}}(\rho)$ in SNM as functions of the reduced density $\rho/\rho_0$. They are also tabulated with their corresponding 90\% and 68\% confidence ranges in Table \ref{EsymTable} and Table \ref{E0table}, respectively.

In Figure \ref{E0Esym}, the black, blue and red shadows represent the prior, 90\% and 68\% credible areas, respectively. As mentioned in the introduction,
both the $E_{\mathrm{sym}}(\rho)$ and $E_{\mathrm{0}}(\rho)$ are constrained into narrow bands  at densities smaller than about 2.5$\rho_0$. However, the symmetry energy diverges broadly at higher densities because of the poor constraint on $J_{\mathrm{sym}}$. More quantitatively,  the symmetry energy at 2$\rho_0$ is found to be $E_{\mathrm{sym}}(2\rho_0)$ =39.2$_{-8.2}^{+12.1}$ MeV at 68\% confidence level as shown in Table \ref{EsymTable}. This value is consistent with the values extracted very recently from several other studies albeit not always with quantified uncertainties. For example, $E_{\mathrm{sym}}(2\rho_0)=46.9 \pm$10.1 MeV at 100\% confidence level was found in ref. \citep{Zhang19a} by inverting several NS observables in the 3-dimensional $J_{\mathrm{0}}$-$K_{\mathrm{sym}}$-$J_{\mathrm{sym}}$ high-density EOS parameter space while all other low-order parameters are fixed at their currently known most probable vales. While this comparison is somewhat unfair because of the different assumptions and methods used, the general agreement is encouraging. Interestingly, an upper bound of 53.2 MeV was derived very recently in ref. \citep{PKU-Meng} by studying the radii of neutron drops using the state-of-the-art nuclear energy density functional theories. In addition, $E_{\rm sym}(2\rho_0)\in[39.4_{+7.5}^{-6.4}, 54.5_{+3.1}^{-3.2}]$ MeV was found from combined analyses of several NS observables and terrestrial laboratory data \citep{LWChen19}. The value of $E_{\mathrm{sym}}(2\rho_0)$ we inferred also falls well within the range of about 26$\sim$66 MeV from the Bayesian analyses of ref. \citep{Baillot19}. Moreover, in a recent study of cooling timescales of protoneutron stars \citep{Nakazato19},  the $E_{\rm sym}(2\rho_0)$ was used as a controlling parameter in constructing a series of phenomenological EOS models. It was found that EOS modes with $E_{\mathrm{sym}}(2\rho_0)$=40$\sim$60 MeV can account for the NS radius and tidal deformability indicated by GW170817. Overall, results from all of these studies are in general agreement within the still relatively large uncertainties. 

At this point, it is useful to note that constraining the $E_{\mathrm{sym}}(\rho)$ at supra-saturation densities has been a longstanding goal of the low-intermediate energy heavy-ion reaction community. However, current results on the $E_{\mathrm{sym}}(2\rho_0)$ based on existing data from heavy-ion reactions with stable beams are still inconclusive \citep{BALI19,Wolfgang}. Quantitatively, the 68\% confidence upper boundary of $E_{\mathrm{sym}}(2\rho_0)$ inferred in this work overlaps with the 100\% confidence lower boundary of  the $E_{\mathrm{sym}}(2\rho_0)$  extracted by the ASY-EOS Collaboration from analyzing their data on the relative flows of neutrons with respect to (w.r.t) protons, tritons w.r.t.$^3$He and yield ratios of light isobars \citep{russ11,ASY-EOS}. Our results, however, is significantly above the super-soft $E_{\mathrm{sym}}(2\rho_0)$ preferred by some transport model analyses of the earlier data on the charged pion ratio \citep{XiaoPRL}. Hopefully, ongoing and planned new experiments especially with high-energy radioactive beams at several advanced rare isotope beam facilities together with improved analysis tools more systematically tested by the broad reaction community will provide more rigorous terrestrial constraints on the $E_{\mathrm{sym}}(\rho)$ at supra-saturation densities. Then, a much more meaningful comparison of the high-density nuclear symmetry energy functionals extracted from astrophysical observations and terrestrial experiments can be made.

As discussed earlier, the $J_0$ parameter has the strongest influence on the maximum mass of neutron stars. Because of the sharp cut-off at 1.97 M$_{\odot}$ used in the default calculations, the PDF of $J_0$ is rather focused. Consequently, as shown in the lower panel of Figure \ref{E0Esym}, the $E_{\mathrm{0}}(\rho)$ in SNM is well constrained up to about $4\rho_0$.  More quantitatively, $E_0(4\rho_0)$=63.3$_{-6.6}^{+19.7}$ MeV at 68\% confidence level, as shown in Table \ref{E0table}. In a recent study \citep{Zhang19apj},  influences of the recently reported mass $M$=2.17$_{-0.10}^{+0.11}$ M$_{\odot}$ of PSR J0740+6620 on the EOS of neutron-rich nuclear matter were analysed. It was found that this new maximum mass of NSs can raise the lower limit of $J_0$ from about -220 MeV to -150 MeV. It provides a tighter constrain on the EOS of neutron-rich nucleonic matter, especially its symmetric part $E_0(\rho)$. This point is verified by the present work as we shall discuss in the next section.
\begin{figure}[ht]
\begin{center}
  \includegraphics[width=13cm,height=13cm]{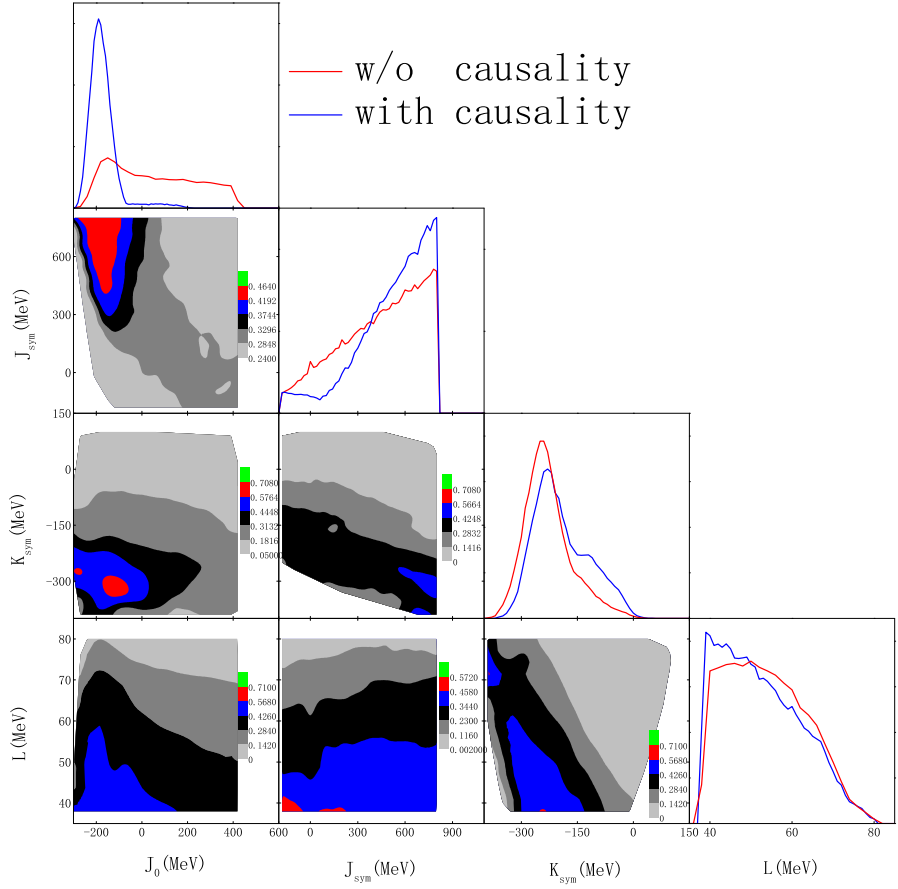}
  \caption{Posterior probability distribution functions of high-density EOS parameters with and without requiring the causality condition.
  The correlation contours are for the case without enforcing the causality condition.
}\label{CausalityCut}
\end{center}
\end{figure}

\section{Individual roles of key model ingredients and neutron star data in Bayesian inference of the high-density EOS parameters}
As we discussed earlier, different ways of parameterizing the EOS have been used in the literature. In particular, not all isospin-dependent parameterizations use simultaneously both the $J_0$ and $J_{\mathrm{sym}}$ terms. How does this different treatment affects what we extract from the same data the high-density behavior of nuclear EOS? Thanks to the continuing great efforts in astronomy, a new record of the maximum mass of NSs may be set at anytime. How does the variation of the NS maximum mass affect what we extract from the same radius data the high-density behavior of nuclear EOS? In addition, it has been pointed out that whether one use a sharp-cut off or a Gaussian function for the NS maximum mass may affect significantly what one can infer from the Bayesian analyses \citep{Miller19}. How does this affect what we learn about the high-density behavior of nuclear EOS? Moreover, we have a dream: the $R_{1.4}$ of canonical NSs has just been measured to better than 5\% $1\sigma$ statistical accuracy with no disagreement from different observations and/or analyses. How does this dreamed case help improve our knowledge about the EOS compared to the default calculations in the previous section?
In the following, we discuss results of our studies on these questions.

\subsection{The role of causality condition}
The causality condition requires that the speed of sound, defined by $v_s^2=dP/d\epsilon$, should not exceed the speed of light, namely, $0\leq v_s^2\leq c^2$.
This condition naturally limits some of the EOS parameters through the pressure $P$ and the energy density $\epsilon$.
While to our best knowledge there is no disagreement about whether the causality condition should be enforced or not, it is educational and interesting to compare results obtained with and without enforcing the causality condition. At least, it allows us to check if our Bayesian inference code does what it is supposed to do. Moreover, we can learn which parameters are mostly affected by the causality condition.

Shown in Figure \ref{CausalityCut} are the posterior PDFs of high-density EOS parameters with or without considering the causality condition. The correlation contours are obtained without using the causality condition. Compared to the PDFs obtained with causality, a wider range extending to large positive values of $J_0$ with higher probabilities is now allowed. This is simply because the causality condition sets the absolute upper limit of the NS maximum mass and consequently the upper limit of $J_0$. Without enforcing the causality condition, the most stringent constraint on $J_0$ now comes from the requirement to support NSs at least as massive as 1.97 M$_{\odot}$. However, this only sets a lower limit for $J_0$ while its upper limit has been removed. Because of its anti-correlation with $J_0$, the parameter $J_{\mathrm{sym}}$ now has higher probabilities to be small when the $J_0$ has higher probabilities to be large. It is also worth noting that the small peak for the PDF of $K_{\mathrm{sym}}$ at about -100 MeV disappears now. Different from the correlations among $J_{\mathrm{sym}}$, $K_{\mathrm{sym}}$ and $J_0$ shown in Figure \ref{pdf-cor1}, due to the larger and almost the same probability for $J_0$ in the range of 0 MeV $\leq J_0 \leq$ 400 MeV, the corresponding $J_{\mathrm{sym}}$ has the same probability in the range of -200 MeV $\leq J_0 \leq$ 0 MeV. This makes the small shoulder in the PDF of $K_{\mathrm{sym}}$ disappear when the causality condition is switched off. As shown by the blue curves in Figure \ref{CausalityCut}, both the $J_{\mathrm{sym}}$ and $K_{\mathrm{sym}}$ obtain higher probabilities to have higher values when the causality condition is turned on. Consequently, the high-density symmetry energy becomes stiffer when the causality condition is enforced. We also notice that the causality as a high-density condition has little effect on the PDFs of $L$ and the low-order parameters as one expects.

\begin{figure*}[htb]
\begin{center}
\resizebox{0.7\textwidth}{!}{
  \includegraphics{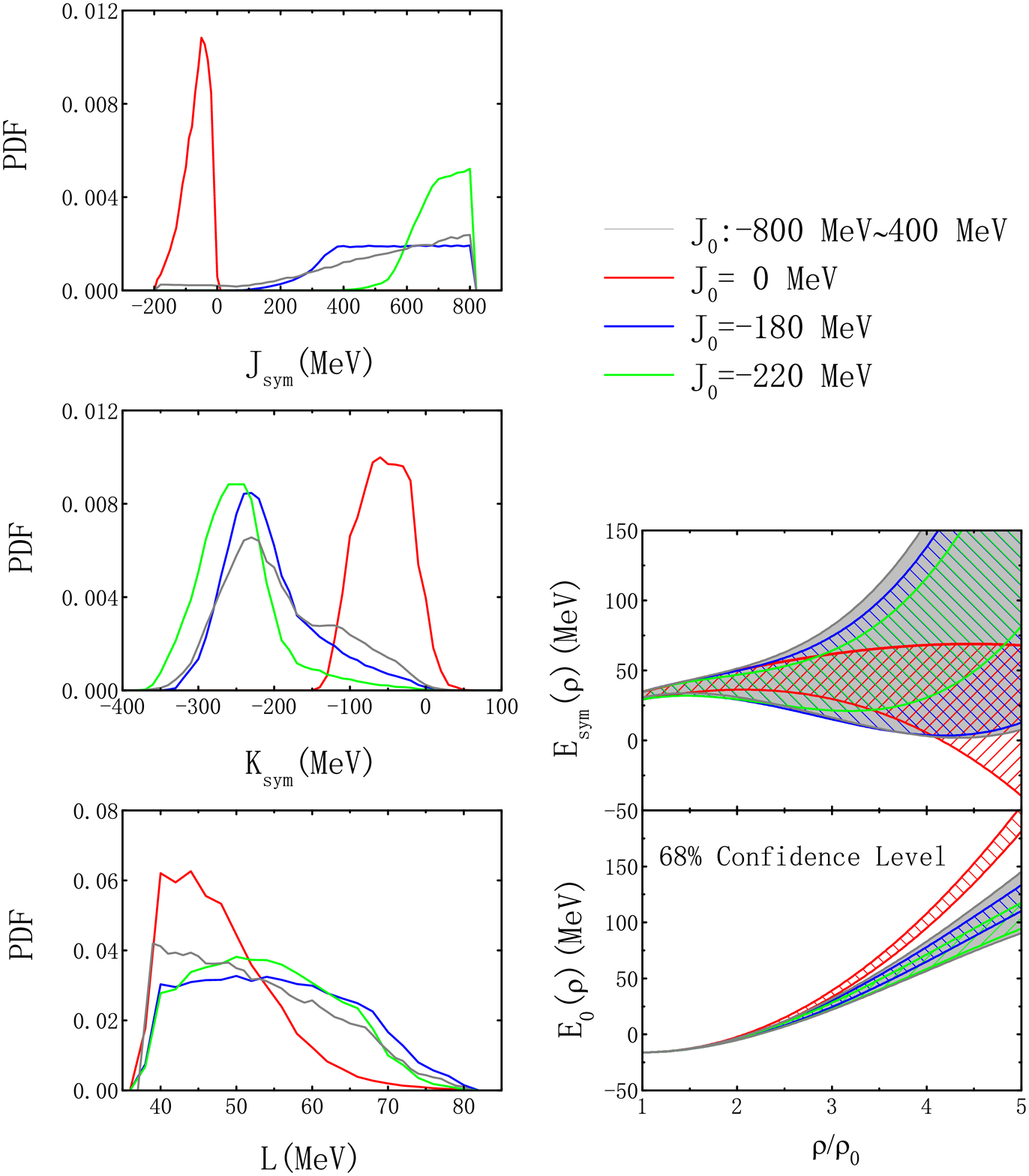}
  }
  \caption{Left: Posterior probability distribution functions of high-density EOS parameters by setting the parameter $J_0$ to 0, -180 and -220 MeV, respectively.
  Right: The role of $J_0$ parameter on inferring the high-density symmetry energy and nucleon specific energy $E_{\mathrm{0}}(\rho)$ in SNM.
   }\label{J0Effect}
\end{center}
\end{figure*}

\subsection{The role of $J_0$ parameter of high-density SNM EOS}\label{E-J0}
The high-order parameter $J_0$ characterizes the high-density behavior of the SNM EOS $E_{\mathrm{0}}(\rho)$. To examine its role, shown in the left panels of Figure \ref{J0Effect} are the PDFs of $J_{\mathrm{sym}}$, $K_{\mathrm{sym}}$ and $L$ when the parameter $J_0$ is fixed at 0, -180 and -220 MeV, respectively.
For comparisons, the PDFs (grey lines) for them in the default case with $J_0$ in the range of -800 to 400 MeV shown in Figure \ref{pdf-cor1} are also included.
It is seen that the $J_0$ has significant effects on the PDFs of $J_{\mathrm{sym}}$ and $K_{\mathrm{sym}}$, but less influences on the PDF of $L$.
The effects of $J_0$ on the parameters $K_0$ and $E_{\mathrm{sym}}(\rho_0)$ are very weak and thus not shown here.
As shown in Figure \ref{pdf-cor1}, because of the negative (positive) correlations between $J_0$ and $J_{\mathrm{sym}}$ ($K_{\mathrm{sym}}$), an increase of $J_0$
leads to a decrease of $J_{\mathrm{sym}}$ but an increase of $K_{\mathrm{sym}}$.  It is interesting to note that the $J_{\mathrm{sym}}$ becomes completely negative when the $J_0$ is set to zero. This is understandable by noticing that the most probable vale of $J_0$ is -190 MeV in the default case as shown in Table \ref{SixPara}. Artificially setting $J_0$=0 is equivalent to making the SNM EOS super-stiff with a much higher contribution to the total pressure. Then, to meet the same conditions especially the maximum mass and causality constraints, the contribution to the pressure from the symmetry energy has to be significantly reduced by making the $J_{\mathrm{sym}}$ largely negative to cancel out the increase of the pressure due to the effectively large increase in $J_0$. On the other hand, it is known that the radius of canonical NSs is almost independent of  $J_{\mathrm{sym}}$, while to maintain the same radius the $K_{\mathrm{sym}}$ and $L$ have to be anti-correlated as shown in Fig. 7 of ref. \citep{Zhang19jpg}. It is thus also easy to understand that while the peak of the PDF of $K_{\mathrm{sym}}$ is shifted from about -230 MeV to -50 MeV, the PDF of L shifts appreciably towards smaller L values.

Overall, when the NS radii data and the maximum mass are used to infer nuclear symmetry energy,  whether to keep a $J_0$ term and what is its uncertainty range all play particularly important roles in determining especially the PDFs of $J_{\mathrm{sym}}$ and $K_{\mathrm{sym}}$, namely the high-density behavior of nuclear symmetry energy.
Shown in the right panels of Figure \ref{J0Effect} are the 68\% CFL boundaries of the symmetry energy and the nucleon specific energy $E_{\mathrm{0}}(\rho)$ in SNM as a function of reduced density with $J_0$=0, -180, -220 MeV, respectively. For comparisons, the default results from Figure \ref{E0Esym} are also included. The following observations can be made: (1) The symmetry energy obtained with a constant of $J_0$=-180 MeV (which is very close to the most probable value of -190 MeV in the default case as shown in Table \ref{SixPara}) is close to the one in the default calculation using $J_0$= -800 MeV$\sim$400 MeV; (2) Below about 2.5$\rho_0$, except the case with $J_0$=0 MeV, both the $E_{\mathrm{sym}}(\rho)$ and $E_{0}(\rho)$ are almost independent of the $J_0$ value; (3) The value of $J_0$ is important for understanding the EOS at densities higher than about 3$\rho_0$.

\subsection{The role of $J_{\mathrm{sym}}$ parameter of high-density symmetry energy}\label{E-Jsym}
The parameter $J_{\mathrm{sym}}$ describes the behavior of nuclear symmetry energy at densities higher than about 2$\rho_0$, as shown in the right panels of Figure \ref{xfraction}. As noted above, the parameter $J_{\mathrm{sym}}$ is less constrained by the present $R_{1.4}$ radius data because the average density reached in canonical neutron stars are not so high, and it is known that the radius of canonical NSs is most sensitive to the pressure around 2$\rho_0$. As mentioned earlier, in some studies for various purposes in the literature,  the parameter $J_{\mathrm{sym}}$ is often set to zero, namely, parameterizing the symmetry energy up to $\rho^2$ only.
Therefore, it is useful to study effects of varying $J_{\mathrm{sym}}$ on the PDFs of high-density EOS parameters within the framework of this work. As we shall see next, effects of $J_{\mathrm{sym}}$ can be explained similarly as those of $J_0$ using the same correlations among the high-density EOS parameters.
\begin{figure*}[htb]
\begin{center}
\resizebox{0.7\textwidth}{!}{
  \includegraphics{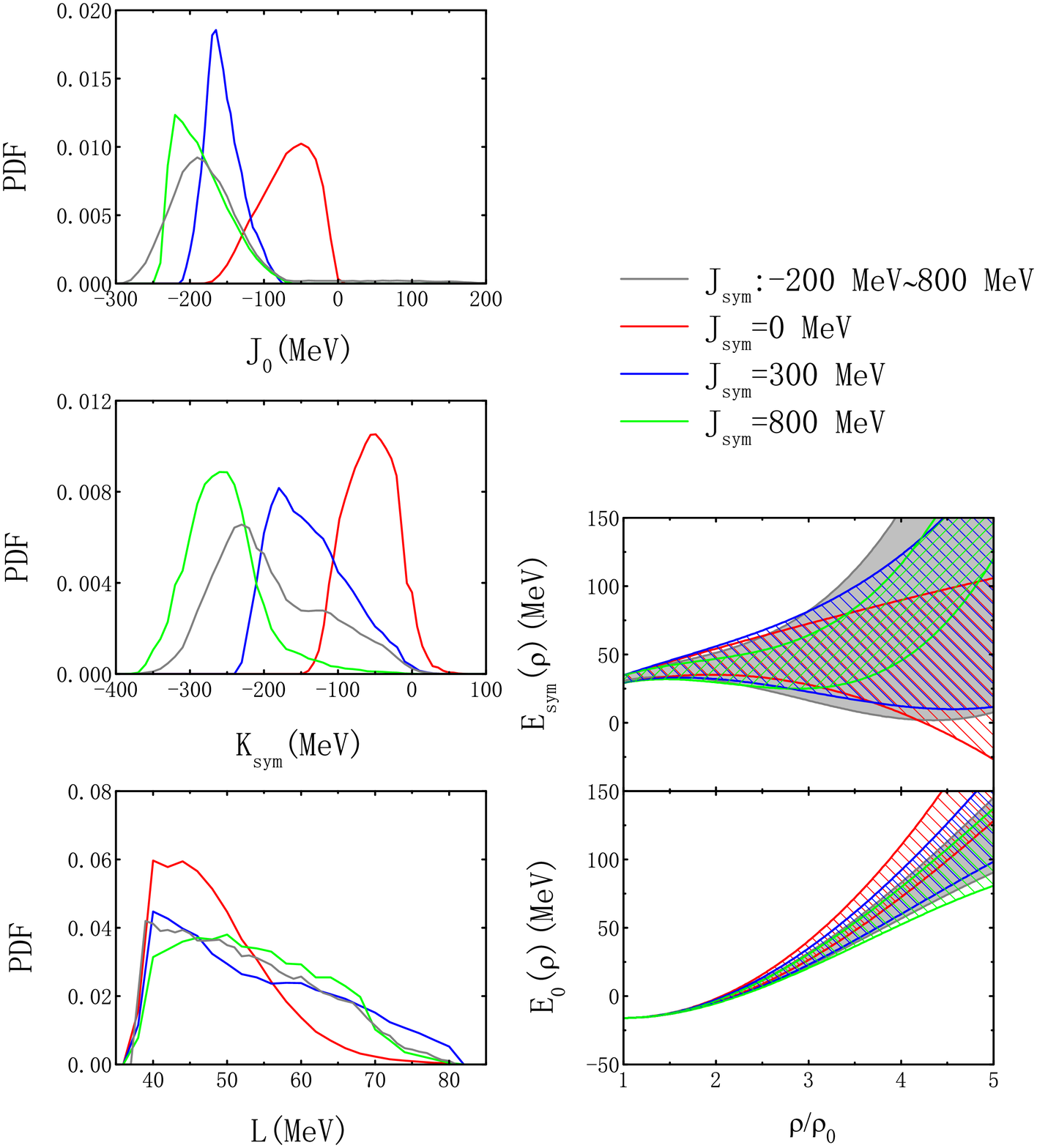}
  }
  \caption{\label{Jsym00} Left: Posterior probability distribution functions of high-density EOS parameters by fixing $J_{\mathrm{sym}}$ at 0, 300 and 800 MeV, respectively. Right: The corresponding effects on inferring the high-density symmetry energy and nucleon specific energy $E_{\mathrm{0}}(\rho)$ in SNM.}
\end{center}
\end{figure*}

Shown in the left panels of Figure \ref{Jsym00} are the PDFs for $J_0$, $K_{\mathrm{sym}}$ and $L$ with $J_{\mathrm{sym}}$=0, 300, 800 MeV, respectively. For comparisons, the PDFs for them from the default calculation (grey lines) using $J_{\mathrm{sym}}$ between -200 MeV and 800 MeV are also included. The PDFs of $K_0$ and $E_{\mathrm{sym}}(\rho_0)$ are not given here because of their weaker correlations with $J_{\mathrm{sym}}$. It is seen that the $J_0$ is always negative when positive values of $J_{\mathrm{sym}}$ are taken. It implies that the contribution to the pressure from the $E_{\mathrm{sym}}\delta^{2}$ term in Eq. (\ref{pre-npe}) with positive $J_{\mathrm{sym}}$ values is high enough that the required contribution from the SNM EOS $E_0(\rho)$ term is smaller. Both $J_0$ and $K_{\mathrm{sym}}$ become larger as $J_{\mathrm{sym}}$ decreases because of the negative correlations between $J_{\mathrm{sym}}$ and $J_0$, as well as between $J_{\mathrm{sym}}$ and $K_{\mathrm{sym}}$. While the variation of $J_{\mathrm{sym}}$ has less influences on the PDF of $L$.

We notice again that the most probable values of $J_0$ and $J_{\mathrm{sym}}$ are -190 MeV and 800 MeV, respectively, in the default calculation. Setting $J_{\mathrm{sym}}$ to zero requires a significant increase of $J_0$, leading to a much stiffer EOS for SNM at densities higher than about 2.5$\rho_0$. This is clearly seen in the lower right panel of Figure \ref{Jsym00} where the 68\% CFL boundary of SNM EOS with the fixed $J_{\mathrm{sym}}$ values are shown. The default results from Figure \ref{E0Esym} are also shown for comparisons. The EOS of SNM becomes softer at densities higher than 2.5$\rho_0$ as $J_{\mathrm{sym}}$ increases. This is due to the fact that $J_0$ becomes small with increasing values of $J_{\mathrm{sym}}$, while the $K_0$ is weakly correlated to $J_{\mathrm{sym}}$. Except the extreme case with $J_{\mathrm{sym}}$ = 800 MeV, both the $E_{\mathrm{sym}}(\rho)$ and $E_0(\rho)$ are almost independent of $J_{\mathrm{sym}}$ at densities below about 2.5$\rho_0$. For the $E_{\mathrm{sym}}(\rho)$,  
the default calculation and the calculations with the fixed $J_{\mathrm{sym}}$ values largely overlap up to about 5$\rho_0$. In particular, the symmetry energy at high densities does not become much stiffer when the fixed value of $J_{\mathrm{sym}}$ increases. This is because the value of $K_{\mathrm{sym}}$ automatically becomes smaller as $J_{\mathrm{sym}}$ increases due to their anti-correlation.

\begin{figure}[ht]
\begin{center}
  \includegraphics[width=13cm,height=13cm]{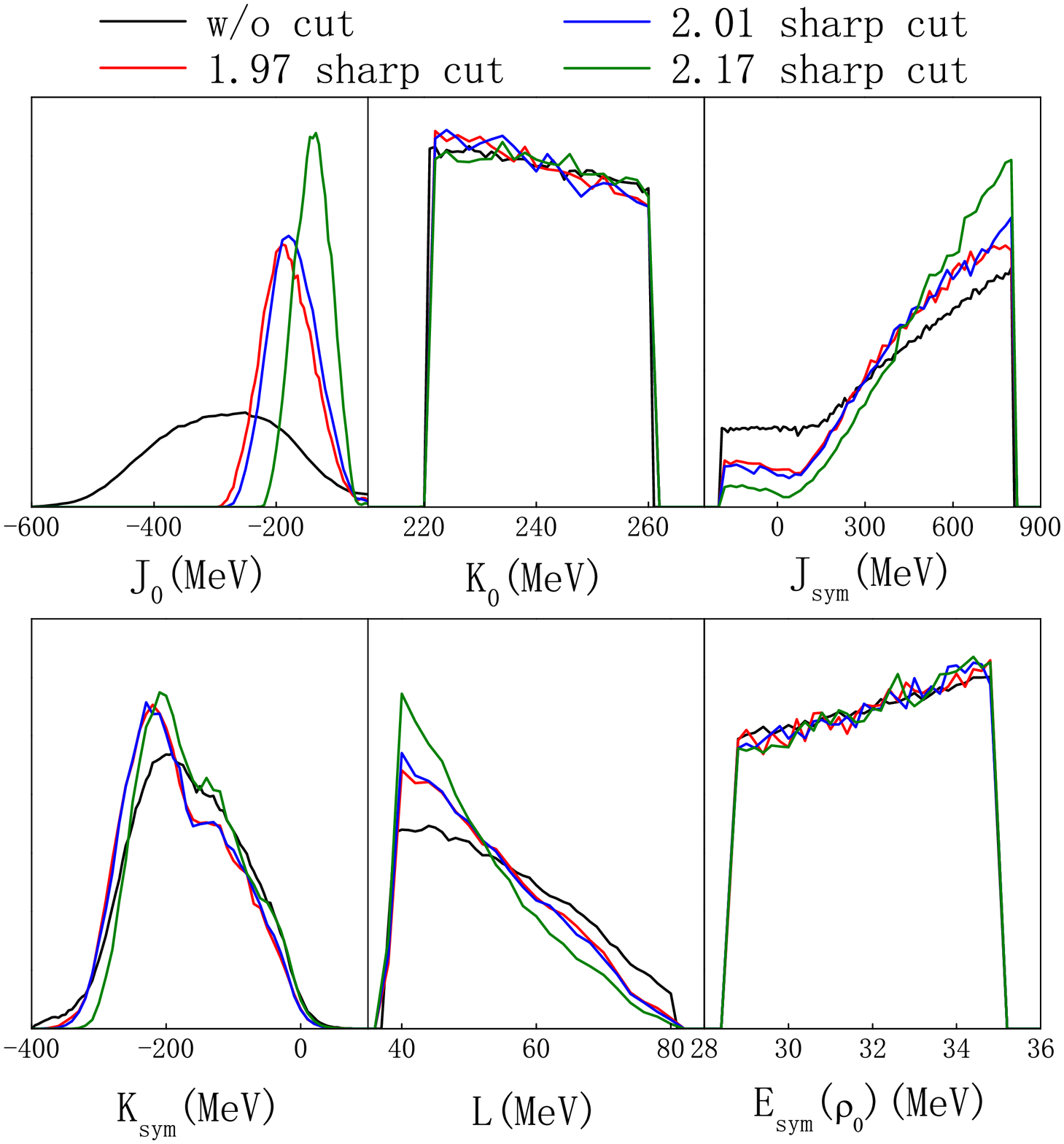}
  \caption{Posterior probability distribution functions of EOS parameters from calculations without a lower limit and with a sharp cut-off at 1.97, 2.01 and 2.17 M$_\odot$ for the NS maximum mass, respectively.}\label{masscut}
\end{center}
\end{figure}

More comments on the case with $J_{\mathrm{sym}}$=0 are necessary. This option has been used in many studies in the literature, see, e.g. refs. \citep{Alam14,Baillot19}. The PDFs for $J_0$, $K_{\mathrm{sym}}$ and $L$ as well as their correlations in this case have been shown already in Figure \ref{Jsym0Cor}. As we noticed earlier, setting $J_{\mathrm{sym}}$=0 leads to the anti-correlation between $J_0$ and $K_{\mathrm{sym}}$ consistent with the result in ref. \citep{Baillot19} but contrary to our default results shown in Figure \ref{pdf-cor1}. This difference has some important consequences. For example, the value of $J_0$= -50$_{-80}^{+40}$ MeV at 68\% CFL is inferred from the present study while $J_0$= 318$_{-366}^{+673}$ MeV was obtained in ref. \citep{Baillot19}. On the other hand, the values of $K_{\mathrm{sym}}$= -50$_{-70}^{+40}$ MeV and $L$= 44$_{-6}^{+16}$ MeV at 68\% CFL we inferred by setting $J_{\mathrm{sym}}$=0  are consistent with the results in ref. \citep{Baillot19}.  Consistent with our conclusions about the role of $J_0$ in inferring the EOS of dense neutron-rich matter, while setting $J_0=0$ affects significantly the accuracy of inferring the high-density symmetry energy, setting $J_{\mathrm{sym}}$=0 affects significantly the accuracy of inferring the high-density EOS of SNM, because of the strong anti-correlation between  the $J_0$ and $J_{\mathrm{sym}}$ parameters.

\begin{figure}[ht]
\begin{center}
  \includegraphics[width=13cm,height=13cm]{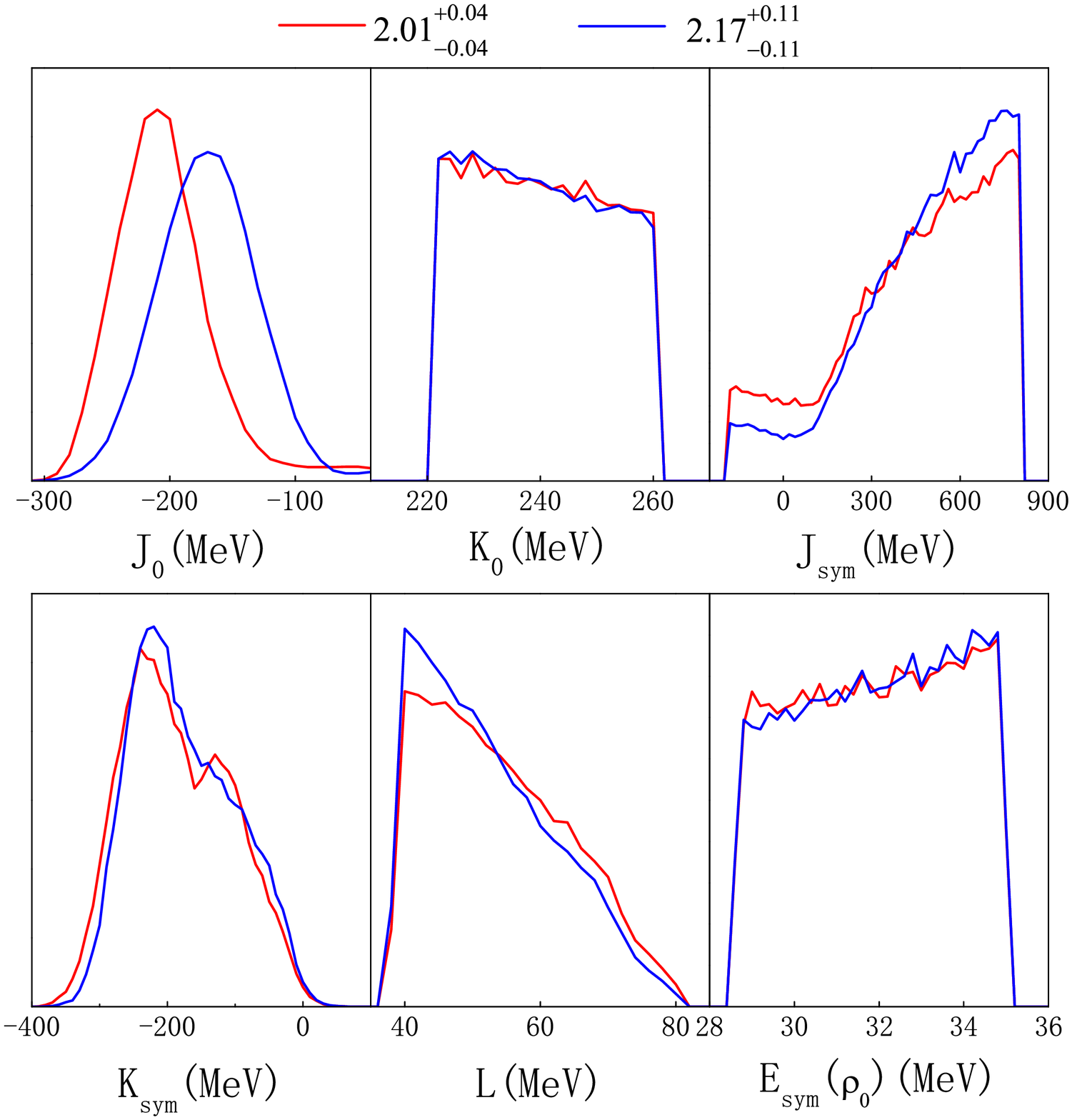}
  \caption{Posterior probability distribution functions of EOS parameters from calculations with a Gaussian distribution centered at 2.01 and 2.17 M$_\odot$, respectively, for the NS maximum mass.}\label{massgaucut}
\end{center}
\end{figure}

\begin{figure*}[htb]
\begin{center}
\resizebox{0.48\textwidth}{!}{
  \includegraphics{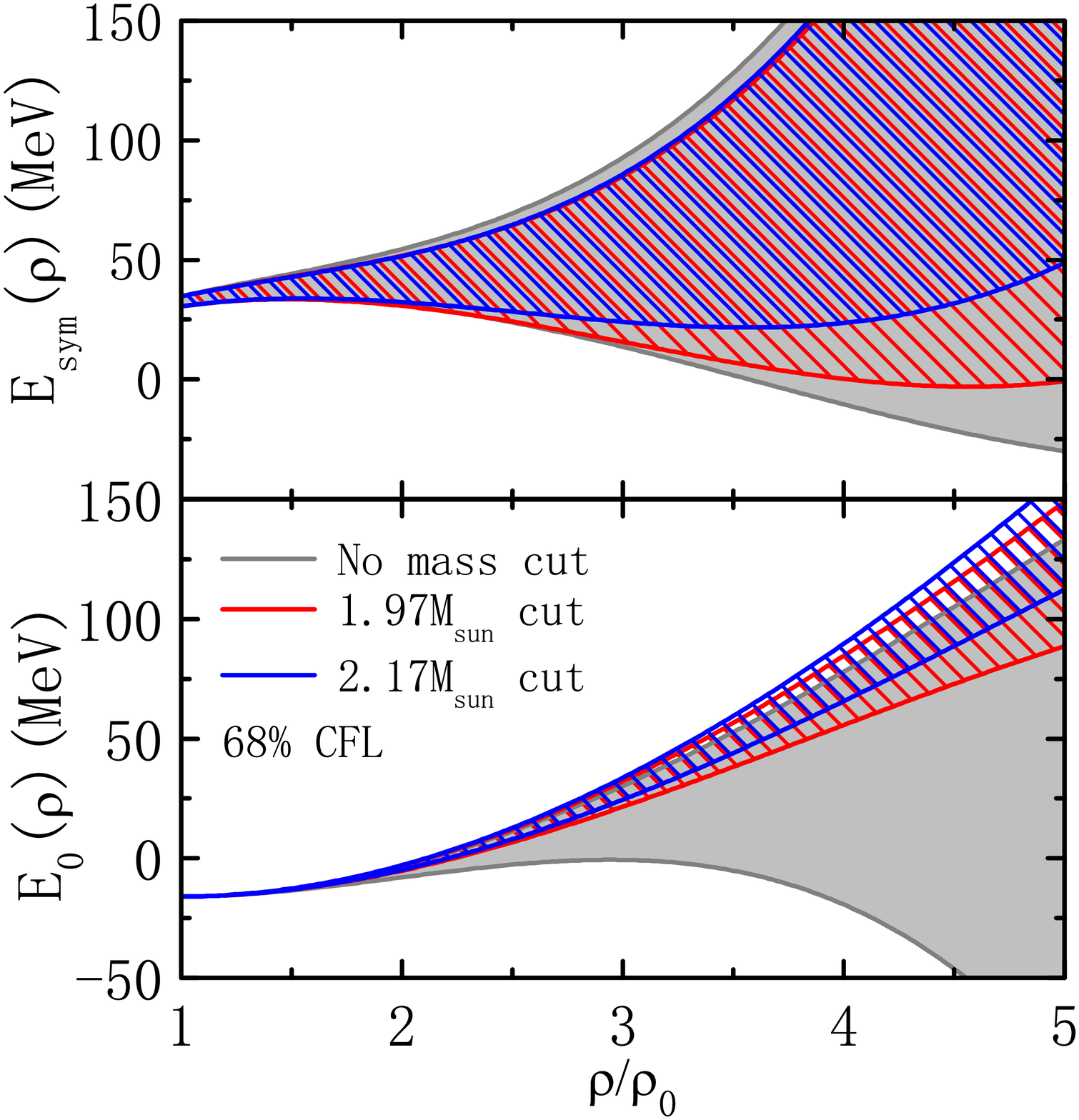}
  }
\resizebox{0.48\textwidth}{!}{
  \includegraphics{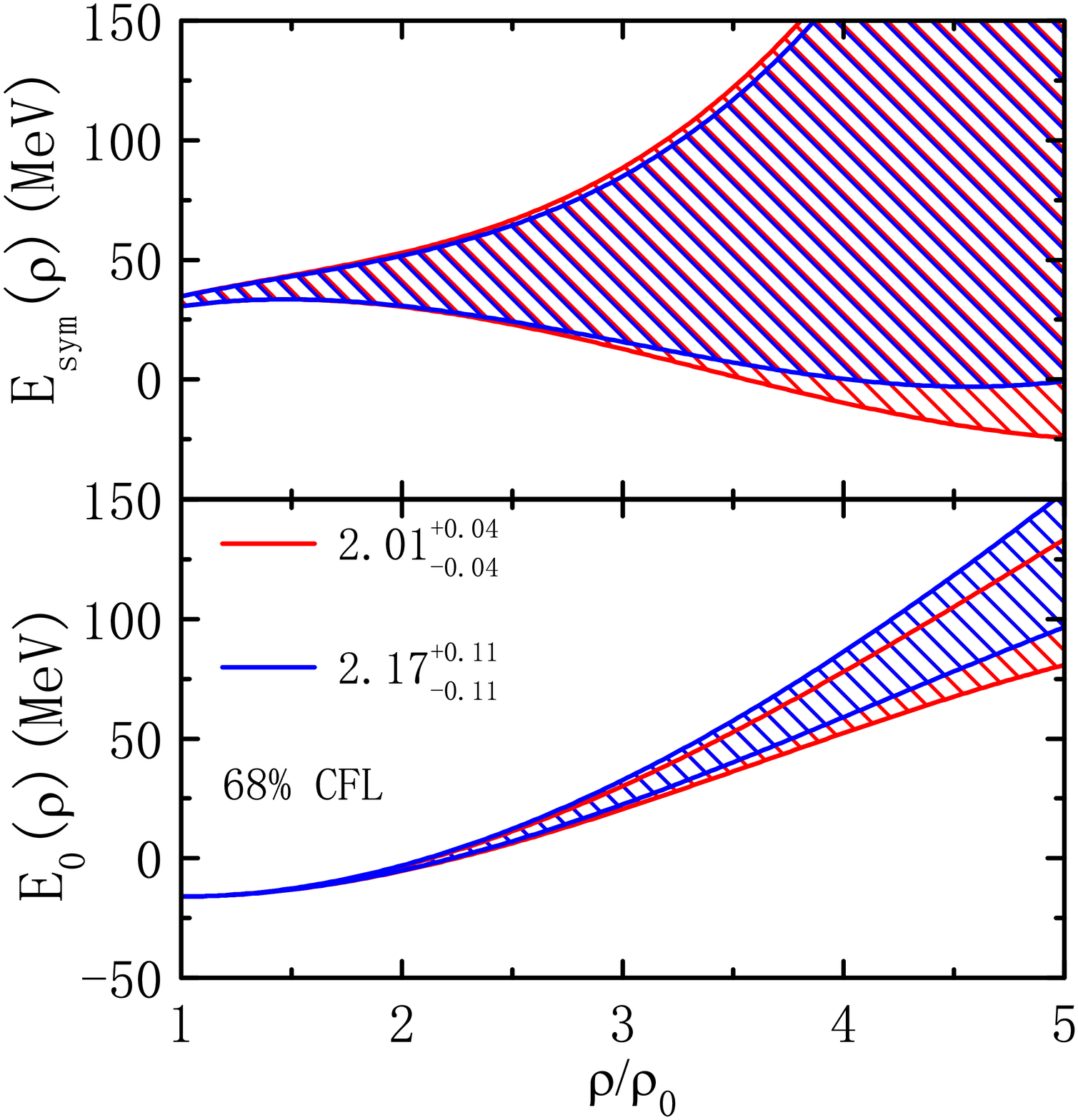}
  }
  \caption{\label{MassE} Effects of using different NS minimum maximum masses on the high-density symmetry energy and nucleon specific energy $E_{\mathrm{0}}(\rho)$ in SNM at 68\% CFL.}
\end{center}
\end{figure*}
\subsection{The role of the observed maximum mass of neutron stars in constraining the high-density EOS}
As mentioned earlier, the maximum mass of NSs is considered as an observable in this work. Since the maximum mass is still changing as more extensive and accurate observations are being made, it is interesting to examine how the value and the way the NS maximum mass information is used may affect what we learn about the high-density EOS from Bayesian inferences. For this purpose, we compare results of the following calculations with the default results: (i) Without requiring a lower limit for the maximum mass (referred as the NS minimum maximum mass in the following) the EOSs have to support; (ii) With a sharp cut-off at 1.97, 2.01 and 2.17 M$_{\odot}$, respectively, for the minimum maximum mass; (iii) The minimum maximum mass is a Gaussian distribution centered at 2.01 and 2.17 M$_{\odot}$ with a $1\sigma$ width of 0.04 M$_{\odot}$ and 0.11 M$_{\odot}$, respectively.

Shown in Figure \ref{masscut} are the PDFs of the EOS parameters from the cases (i) and (ii). Obviously and easily understood, the lower limit of $J_0$ has to go up to support gradually more massive NSs while its upper limit remains the same. Consequently, the most probable value of $J_0$ increases, leading to more stiff SNM EOS as shown in Figure \ref{MassE}. It is interesting to see that the PDF of $J_{\mathrm{sym}}$ shifts to favor higher $J_{\mathrm{sym}}$ values, leading to an increased mean value of the latter when the NS minimum maximum mass increases from 1.97 to 2.17 M$_{\odot}$. Namely, now the mean values of both $J_0$ and $J_{\mathrm{sym}}$ increase seemingly in contradiction with their anti-correlation observed in the default calculation where the NS minimum maximum mass is fixed at 1.97 M$_{\odot}$.  This is easily understandable because the maximum pressure is no longer the same in calculations with different NS minimum maximum masses. Contributions to the increased pressure necessary to support more massive NSs can come from both the high-density SNM EOS and symmetry energy. As a result, the mean values of both $J_0$ and $J_{\mathrm{sym}}$ increase as the NS minimum maximum mass increases, while effects on other EOS parameters are weak. Thus, as demonstrated recently in ref. \citep{Zhang19apj}, more precise measurements of the NS maximum mass will improve our knowledge about the high-density behavior of both SNM EOS and nuclear symmetry energy. The same phenomena are observed when the Gaussian distributions are used for the minimum maximum mass as shown in Figure \ref{massgaucut} for the case (iii). As indicated, one minimum maximum mass distribution centers at 2.01 M$_{\odot}$ with a $1\sigma$ width of 0.04 M$_{\odot}$, while the other one centers at 2.17 M$_{\odot}$ with a $1\sigma$ width of 0.11 M$_{\odot}$. The PDFs of both $J_0$ and $J_{\mathrm{sym}}$ shift towards higher $J_0$ and $J_{\mathrm{sym}}$ values as the central mass increases from 2.01 to 2.17 M$_{\odot}$. Again, this is easily understood.

Effects of using different NS minimum maximum masses on the high-density symmetry energy and nucleon specific energy $E_{\mathrm{0}}(\rho)$ in SNM at 68\% CFL
are shown in Figure \ref{MassE}. The following observations can be made: (1) The maximum mass condition plays an increasingly more appreciable role in determining the symmetry energy only at densities higher than about 2.5$\rho_0$. A higher value for the NS minimum maximum mass raises the lower limit for the high-density symmetry energy mainly due to the increased mean value of $J_{\mathrm{sym}}$ as we just discussed above, while the upper limit does not change much; (2) The maximum mass condition has a significant impact on the SNM EOS at $\rho > 2 \rho_0$.
The increased NS maximum mass requires higher values of $J_0$ while the saturation parameters are basically not affected, stiffening the SNM EOS above about $2\rho_0$;
(3) As shown in the right panels of Figure \ref{MassE}, using the two different Gaussian distributions for the NS minimum maximum mass does not cause any obvious change in the high-density EOS compared to the calculations with the sharp cut-offs. Of course, the two different central masses have some obvious and easily understood effects.
We emphasize that this is probably due to the fact that the radius data we used is only for a single NS of mass 1.4 M$_{\odot}$ instead of a group of NSs involving more massive ones.

\begin{figure*}[htb]
\begin{center}
\resizebox{0.8\textwidth}{!}{
  \includegraphics{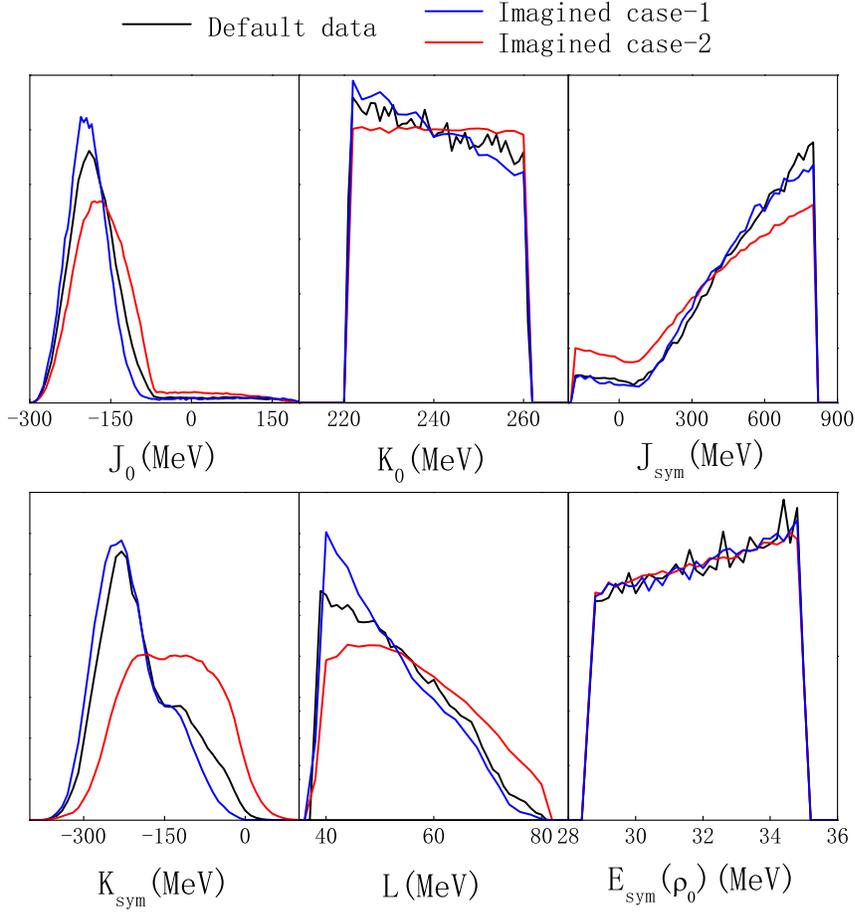}
  }
  \caption{\label{Dream} Effects of different precisions of measuring $R_{1.4}$ shown in Table \ref{tab:data} on the PDFs of EOS parameters.}
\end{center}
\end{figure*}

\subsection{The role of $R_{1.4}$ measurement accuracy on symmetry energy parameters}\label{R14}
Here we study how much better we can infer the EOS parameters by using the two sets of  imaginary data given in Table 1.
The imaginary case-1 containts three data points with mean radii different from each other by about 1 km (i.e., a 10\% systematic error) but the same absolute error bar of 0.8 km at 90\% CFL. The average radius of this case is 11.5 km. While the imaginary case-2 has a single radius of 11.9 km and the same statistical error bar of 0.8 km as in the case-1. The PDFs of EOS parameters from these two imaginary cases are compared with those from our default calculations in Figure \ref{Dream}. It is seen that the case-1 and our default calculations give essentially the same PDFs for all six EOS parameters. Although in case-1, the statistical error is smaller compared to the default case, since the mean radii of the three points are already different by about 1 km, it is not surprising that the PDFs of case-1 are not much different from the default case. The PDFs of the imaginary case-2 are appreciably different from those of the default case, especially for $K_{\mathrm{sym}}$ and $L$. More quantitatively, as listed in Table \ref{SixPara-i}, the most probable values of both $K_{\mathrm{sym}} $ and $L$
increase significantly while their 68\% CFL ranges remain approximately the same as the imaginary case-1 or the default case. In addition, the mean value of $J_0$ slightly increases while the mean value of $J_{\mathrm{sym}}$ decrease slightly.

\begin{table}[tbp]
\centering
\caption{{\protect\small Most probable values of the EOS parameters and their 68\% CFL boundaries obtained using the
imagined data listed in Table 1.}}
\label{SixPara-i}%
\begin{tabular}{ccccccc}
\hline\hline
Quantity &Imagined case-1 &Imagined case-2 \\
\hline
$J_0$ & $-190_{-40}^{+30}$ & $-170_{-50}^{+50}$\\
$K_0$ & $232_{-12}^{+14}$ & $236_{+12}^{-16}$ \\
$J_{\mathrm{sym}}$ & $800_{-340}^{+20}$ & $800_{-440}^{+0}$  \\
$K_{\mathrm{sym}}$ & $-240_{-50}^{+70}$ & $-180_{-40}^{+130}$\\
$L$ & $40_{-2}^{+16}$ & $44_{-6}^{+18}$ \\
$E_{\mathrm{sym}}(\rho_0)$ & $33.8_{-3.2}^{+1.0}$ & $34.6_{-4.0}^{+0.2}$  \\ \hline\hline
\end{tabular}%
\end{table}

Without changing any other conditions, the increase of mean radius from about 11.5 km in the default case to 11.9 km in the imaginary case-2 requires an increase in pressure around $2\rho_0$ while keeping the pressures at high density approximately the same to satisfy the same condition on the maximum mass. The increase in pressure around $2\rho_0$ can be achieved by increasing the $K_0$, $J_0$ and/or $L$ and $K_{\mathrm{sym}}$. While an increase in $J_0$ normally leads to a decrease in $J_{\mathrm{sym}}$ as we discussed earlier due to the maximum mass constraint at high densities. Thus, the observed variations of the PDFs can all be
qualitatively understood. Quantitatively, however, reducing the statistical and removing the systematic error bars in measuring the $R_{1.4}$ does not seem to help narrow down the PDFs of the EOS parameters compared to the default calculation using the real data. While we probably just made up bad numbers in our otherwise very good dream, this finding may be disappointing but not surprising. Our study here using only the radius data of a single canonical NS sets a useful reference. In reality, joint PDFs of the mass-radius measurements are normally inferred from Bayesian analyses of the raw data from observations. A collection of such data extending to heavy mass regions, or the high precision radius data of several NSs with different masses will certainly help put much tighter bounds on the PDFs of the high-density EOS parameters. Our preliminary studies using several different hypothetical mass-radius correlations between 1.2 to 2  M$_{\odot}$ indicate that they lead to very different PDFs of EOS parameters. Results of this study will be reported elsewhere.

\section{Summary and Outlook}
\label{sec4}
In summary, using an explicitly isospin-dependent parametric EOS of nucleonic matter we carried out a Bayesian inference of high-density nuclear symmetry energy  $E_{\rm{sym}}(\rho)$ and the associated nucleon specific energy $E_0(\rho)$ in SNM using the latest $R_{1.4}$ radius data available in the literature, under several general conditions required for all NS models.
The most important physics findings from this study are
\begin{itemize}

\item The available astrophysical data can already improve significantly our knowledge about the $E_0(\rho)$ and $E_{\rm{sym}}(\rho)$ in the density range of $\rho_0-2.5\rho_0$ compared to what we currently know about them based mostly on terrestrial nuclear experiments and predictions of nuclear many-body theories.
In particular, the symmetry energy at 2$\rho_0$ is determined to be $E_{\mathrm{sym}}(2\rho_0)$ =39.2$_{-8.2}^{+12.1}$ MeV at 68\% CFL approximately independent
of the EOS parameterizations used and uncertainties of the absolutely maximum mass of NSs. However, at higher densities, the 68\% confidence boundaries for both the $E_0(\rho)$ and $E_{\rm{sym}}(\rho)$ diverge depending strongly on the EOS parameterizations used and several uncertainties. 

\item A precise measurement of $R_{1.4}$ alone with less than 5\% 1$\sigma$ statistical error and no systematic error will not improve much the constraint on the EOS of neutron-rich nucleonic matter at densities below about $2.5\rho_0$ compared to the constraints extracted from using the available radius data. While we are hopeful that high precision joint PDFs from simultaneous mass-radius measurements extending to heavy mass regions will significantly further narrow down the EOS both below and above $2.5\rho_0$.

\item The radius data and other general conditions, such as the observed NS maximum mass and causality condition introduce strong correlations for the high-order parameters used in parameterizing the $E_0(\rho)$ and $E_{\rm{sym}}(\rho)$. Reflected clearly in the PDFs of the high-density EOS parameters, the high-density behavior of $E_{\rm{sym}}(\rho)$ inferred depends strongly on how the high-density $E_0(\rho)$ is parameterized, and vice versa. This is particularly true for densities higher than about $2.5\rho_0$ where the third-order parameters $J_0$ and $J_{\rm{sym}}$ play the dominating role. Since these two parameters are not always used simultaneously in parameterizing the $E_0(\rho)$ and $E_{\rm{sym}}(\rho)$ in the literature, different correlations among the PDFs of EOS parameters and thus different high-density behaviors of the symmetry energy may be inferred from the same set of NS observational data.

\item The value of the observed NS maximum mass and whether it is used as a sharp cut-off for the minimum maximum mass or through a Gaussian distribution in the Bayesian analyses affect significantly the lower boundaries of $E_0(\rho)$ and $E_{\rm{sym}}(\rho)$ only at densities higher than about $2.5\rho_0$. While the EOS constraints extracted in the density region of $\rho_0-2.5\rho_0$ are not influenced by the remaining uncertainties about the NS absolutely maximum mass.
\end{itemize}

Finally, we speculate that radii of more massive NSs and additional messengers especially those directly from NS cores or emitted during collisions between two NSs or heavy nuclei will be useful to further constrain the EOS of dense neutron-rich nuclear matter especially at densities higher than about $2.5\rho_0$.

\acknowledgments
We thank Drs. Nai-Bo Zhang and De-Hua Wen for helpful discussions and Dr. Matt Wood for providing us the high-performance computing resources used for carrying out  this work. Wen-Jie Xie is supported in part by the China Scholarship Council and appreciates the productive research conditions provided to him at Texas A\&M University-Commerce.
BAL acknowledges the U.S. Department of Energy, Office of Science, under Award Number DE-SC0013702, the CUSTIPEN (China-U.S. Theory Institute for Physics with Exotic Nuclei) under the US Department of Energy Grant No. DE-SC0009971.

\end{document}